\documentclass[twocolumn]{aastex701}

\usepackage{apjfonts}
\usepackage{makecell}
\usepackage{amsmath}
\usepackage{longtable}
\usepackage{xtab}
\usepackage{pifont}
\usepackage{hyperref}
\usepackage{footnote}
\usepackage{multirow}
\usepackage{threeparttable}
\usepackage{booktabs}

\begin{document}

\title{Two Earliest Optical-UV Tidal Disruption Events Hidden in the SDSS DR7 Catalog Unveiled by the Transformer-Based Spectrum Classifier}

\author[orcid=0009-0006-4622-1417]{Ranfang Zheng}
\affil{Department of Astronomy, University of Science and Technology, Hefei, 230026, China}
\affil{School of Astronomy and Space Sciences, University of Science and Technology of China, Hefei 230026, China}
\email{rfzheng@mail.ustc.edu.cn}

\author[0000-0003-4959-1625]{Zheyu Lin}
\affil{Université Paris Cité, CNRS, Astroparticule et Cosmologie, F-75013 Paris, France}
\affil{Department of Astronomy, University of Science and Technology, Hefei, 230026, China}
\affil{School of Astronomy and Space Sciences, University of Science and Technology of China, Hefei 230026, China}
\email{zheyu.lin@apc.in2p3.fr}

\author[0000-0002-7660-2273]{Xu Kong}
\affil{Department of Astronomy, University of Science and Technology, Hefei, 230026, China}
\affil{School of Astronomy and Space Sciences, University of Science and Technology of China, Hefei 230026, China}
\affil{Institute of Deep Space Sciences, Deep Space Exploration Laboratory, Hefei 230026, China}
\email{xkong@ustc.edu.cn}

\correspondingauthor{Zheyu Lin}
\email{zheyu.lin@apc.in2p3.fr}
\correspondingauthor{Xu Kong}
\email{xkong@ustc.edu.cn}

\author{Ran Song}
\affil{Department of Astronomy, University of Science and Technology, Hefei, 230026, China}
\affil{School of Astronomy and Space Sciences, University of Science and Technology of China, Hefei 230026, China}
\email{ransong@mail.ustc.edu.cn}

\author{Dezheng Meng}
\affil{Department of Astronomy, University of Science and Technology, Hefei, 230026, China}
\affil{School of Astronomy and Space Sciences, University of Science and Technology of China, Hefei 230026, China}
\email{dezhengmeng@mail.ustc.edu.cn}

\author[0000-0003-4200-4432]{Lulu Fan}
\affil{Department of Astronomy, University of Science and Technology, Hefei, 230026, China}
\affil{School of Astronomy and Space Sciences, University of Science and Technology of China, Hefei 230026, China}
\affil{Institute of Deep Space Sciences, Deep Space Exploration Laboratory, Hefei 230026, China}
\affil{College of Physics, Guizhou University, 550025 Guiyang, PR China}
\email{llfan@ustc.edu.cn}

\begin{abstract}
Optical spectroscopic features are decisive in the current identification of optical-UV tidal disruption events (TDEs). Regarding that the TDE estimated occurrence rate is $10^{-5}-10^{-4}$~galaxy$^{-1}$~yr$^{-1}$ by both theoretical and observational methods, large optical spectroscopic catalogs with $>$10$^5$ galaxy spectra can include some serendipitous spectra with TDE spectroscopic features, which can be found after building a useful selection method. 
We hereby introduce a principal component analysis enhanced Transformer TDE spectrum classifier which achieves a precision of 0.88 and a recall of 0.99 on our evaluation dataset, and report its inspiring discoveries in the widely-used SDSS DR7 catalog: two newly discovered TDEs and one reported likely TDE.
For SDSS~J124225.39+642919.0, we confirm the presence of a UV transient in GALEX catalog when the spectrum was taken, and its occurrence time should be earlier than the spectrum observation time, MJD $<$ 52316 (February 11, 2002), making it the earliest optical-UV TDE discovered by now. For SDSS~J152459.70+045423.1, its spectrum matches all features of the TDE-H+He spectrum, and was taken during an optical outburst recorded by the Catalina Real-time Transient Survey. The start of this outburst lies in 54269 $<$ MJD $<$ 54476 (June 18, 2007 - January 11, 2008), making it one of the earliest among the reported optical TDEs. The discovery of two new TDEs highlights the power of machine-learning based classifiers in digging out buried treasures in large-volume catalogs, and marks a new method for discovering optical-UV TDEs. 
\end{abstract}




\section{Introduction} \label{sec:intro}
A tidal disruption event (TDE) occurs when a star passes too close to a black hole (BH) and the tidal force of the BH overcomes the self-gravity of the star \citep{Hills1975, Rees1988}. Early works predict that a TDE should exhibit a transient flare whose luminosity rises in a few months and declines in a few months to years, and the spectral energy distribution (SED) should peak at soft X-ray to far-ultraviolet (FUV) bands \citep[e.g.,][]{Rees1988}. Following these predictions, TDEs were initially discovered in soft X-ray bands in late 1990s \citep{Bade1996, Komossa1999}, and later in UV bands in 2006 \citep{Gezari2006}. 

In details, the first UV TDE, GALEX D3-13, was discovered in one of the most frequently visited fields during the GALEX \citep{Martin2005} Deep Imaging Survey (DIS). 
In the first epoch in 2003, the source was not detected in both NUV (near-UV) and FUV bands, but in all epochs in 2004, the source was detected in both bands. In the next two years, the source became fainter, its luminosity can be well fitted with a power-law decline model. The connection between soft X-ray and UV TDEs was highlighted by a transient soft X-ray counterpart. In two Chandra exposures of equal length, it was detected in April 2005 but not detected in September 2005. Later on, two more UV TDEs GALEX D1-9 and D23H-1 were dug out from the GALEX DIS data \citep{Gezari2008,Gezari2009}. In these two TDEs, optical counterparts were discovered for the first time ever, which were both fainter than UV counterparts, and soft X-ray counterparts were also detected. The soft X-ray, UV and optical counterparts were thus unified by the frame of TDE, which deeply influenced later TDE observational studies. 

\citet{vanVelzen2011} reported two TDEs selected from the archival SDSS Stripe 82 multi-epoch imaging data, which covered a sky area of $\sim$300 deg$^2$. These two TDEs were carefully filtered out by cuts of galaxy-star type, flux difference, imaging quality, on-center position, active galactic nuclei (AGNs) and visual inspection. The fitted peak dates of these two TDEs are MJD = 53974 (August 27, 2006) and 54357 (September 14, 2007). Since 2012, the discovery rate of optical TDEs has entered an upward channel thanks to the involvement of several time-domain wide-field sky surveys, e.g., Pan-STARRS \citep[e.g.,][]{Gezari2012}, ASAS-SN \citep[e.g.,][]{Holoien2014}, ATLAS and (i)PTF/ZTF \citep[e.g.,][]{Arcavi2014, Blagorodnova2017, vanVelzen2021, Hammerstein2023, Yao2023}. In particular, ZTF has lifted the discovery rate from $\sim$2 to $\sim$20 per year, and probably more importantly, the ZTF sample articles \citet{vanVelzen2021} and \citet{Hammerstein2023} set up current optical spectroscopic criteria for TDEs. In TDE spectra, a blue continuum above the host-galaxy spectrum is mandatory, and the presence or absence of the transient broad Balmer and helium emission lines determines that the TDE belongs to which of these four spectroscopic classes: TDE-H, TDE-He, TDE-H+He (Also known as TDE-Bowen since Bowen fluorescence emission lines usually present) and TDE-featureless.

The occurrence rate of TDEs is estimated to be $10^{-5}-10^{-4}$~galaxy$^{-1}$~yr$^{-1}$, by both theoretical \citep[e.g.,][]{Lightman1977, Magorrian1999, Syer1999, Wang2004, Stone2016} and observational methods \citep[e.g.,][]{Donley2002, Esquej2008, Gezari2009, vanVelzen2014, Holoien2016, vanVelzen2018, Yao2023}, and spectroscopic features are keys to identifying TDEs. Inspired by these two facts, we realize that it is totally reasonable that large optical spectroscopic surveys with total galaxy spectrum number of $>$10$^5$ can serendipitously take some galaxy spectra when TDEs are ongoing inside, and we should be able to find and identify them after building a useful selection method. This spectroscopic search should be meaningful. First, the total number of optical-UV TDEs is still less than 200, and under an optimistic assumption of TDE occurrence rate of $10^{-4}$~galaxy$^{-1}$~yr$^{-1}$, we can discover up to hundreds of TDEs in a catalog with $\geqslant$$10^6$ available spectra, e.g., SDSS or DESI. Second, it is still unclear why these spectroscopic features are related to TDEs, making these spectroscopic criteria empirical. If the sample size reaches a few tens, we can perform statistical analysis to examine whether these spectroscopic features only exhibit during TDEs. The machine learning algorithms excel in dealing with large samples, enable large classification programs. Thus, following the successful creation of a Transformer-based photometric TDE classifier (\texttt{TTC}, \citealt{Zheng2025}), we hereby carry out a new project - a Transformer-based TDE spectrum classifier that enhanced by the principal component analysis (PCA), and our aim is to find serendipitously taken TDE-like spectra among large spectroscopic survey catalogs, and judge if these are ascribed to TDEs with the help of photometric data at the time when the spectra were taken. 

In this work, we report the classification results of the SDSS DR7 catalog \citep{SDSSDR7}. Three TDEs are selected, and two of which are newly discovered. For the two newly discovered TDEs, we constrain their occurrence dates to MJD $<$ 52316 (February 11, 2002) and 54269 $<$ MJD $<$ 54476 (June 18, 2007 - January 11, 2008), respectively, making the former 
source undoubtedly the earliest optical-UV TDE discovered by now.
The discovery of these two new TDEs proves the capability of our Transformer-based TDE spectrum classifier, and highlights the potential of machine-learning classifiers in digging out buried treasures in archival large-volume catalogs, even most widely used ones. 

The structure of the paper is as follows. 
In Section~\ref{sec:method}, we briefly describe the classification methodology adopted in this work, together with the underlying principles of the PCA–Transformer architecture. 
In Section~\ref{sec:result}, we list the TDE candidates, and judge if these candidates are reliable TDEs. In Section~\ref{sec:conclusion} we conclude the study with a summary of our findings, and discuss the future application of this TDE spectrum classifier. The magnitudes in this paper are in AB units \citep{Oke1974}.

\section{Method}
\label{sec:method}

We employ the PCA–Transformer model (Section~\ref{sec:model_sc}) to classify spectra that contain a mixture of transient sources and their host galaxies. 
The whole procedure consists of four main components: original data preparation (Section~\ref{sec:data_prep}), evaluation dataset preparation (Section~\ref{sec:evaluation}), model training and evaluation (Section~\ref{sec:Model Training and Evaluating}) and model application (Section~\ref{sec:real_model_tt}). 



\begin{figure*}[htbp]
\centering
\includegraphics[width=1.0\textwidth]{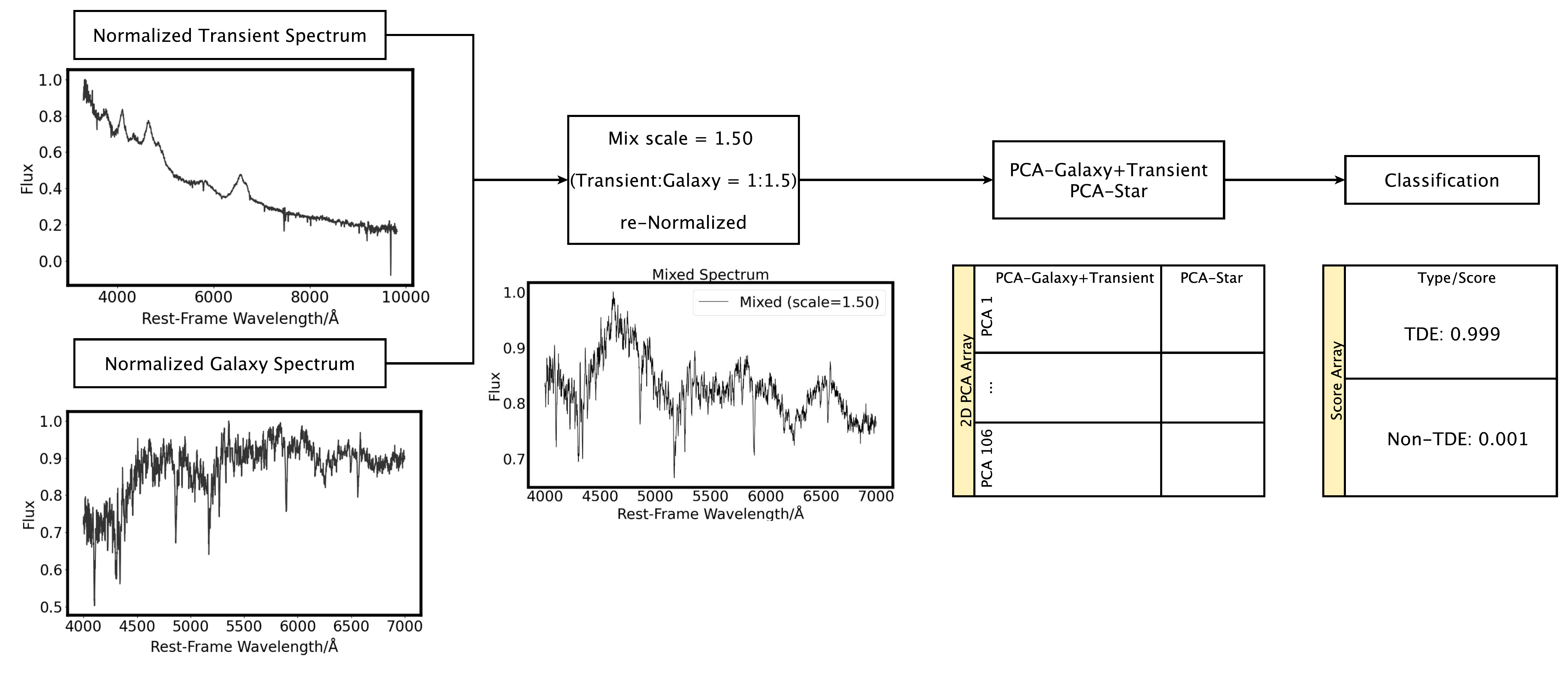}
\caption{Schematic illustration of the preparation procedure for the training and test datasets. We first select spectra of the transient source and its host galaxy. After applying redshift correction, each spectrum is normalized by its maximum flux. The two spectra are then superimposed using different mixing scales, where the mixing scale represents the flux ratio of the host galaxy to the transient source during the combination process. 
The combined spectrum is subsequently re-normalized by its maximum flux. Finally, PCA is performed on the mixed spectrum, and the resulting coefficient matrix is used as the input for training the classification model.
\label{fig:processing_data}}
\end{figure*}

\subsection{PCA-Transformer Structure}
\label{sec:model_sc}

Here, we propose a PCA–Transformer architecture that integrates the dimensionality reduction capability of PCA with the Transformer’s ability to model contextual relationships among PCA coefficients.   Rather than directly classifying the raw spectra, our method performs classification based on the coefficient matrices obtained by processing each target spectrum onto a set of predefined PCA templates.


Our approach is inspired by PCA-Net, a deep learning framework with a simple and interpretable architecture originally developed for image classification \citep{PCA-net1}. In PCA-Net, PCA is used to construct multi-stage filter banks, enabling effective feature extraction. Previous study has shown that this architecture can mitigate the curse of dimensionality \citep{PCA-Net2}.
We consider this approach is particularly effective for spectral data, which are inherently high-dimensional, especially in the case of high-resolution spectra.

The PCA–Transformer framework can be viewed as a two-stage process of dimensionality reduction followed by dimensionality expansion. In the reduction stage, PCA templates are employed to suppress noise-dominated components as effectively as possible. In the expansion stage, the transformer leverages its hidden representations to capture nonlinear correlations among the resulting coefficients.

One reason for adopting this method is that directly using the original spectra incurring substantial GPU computational costs. In practice, the PCA step can be performed on the CPU, after which only the reduced feature representations are processed by the transformer model on the GPU.

Furthermore, when high dimensional spectra are directly used as model inputs, predictive performance can degrade significantly if the inference data differ in quality from the training set, for example in signal-to-noise ratio (SNR), spectral resolution, or instrumental response. Such degradation, commonly attributed to distributional difference between training and application domains, has been reported in previous spectral classification studies (e.g., \citealt{Iascore}). Incorporating PCA for preliminary dimensionality reduction can alleviate this issue by projecting the data onto a low dimensional subspace that captures the dominant variance and global spectral structure. When variations in resolution and signal to noise ratio remain limited, the leading PCA coefficients are generally stable, enabling the classifier to operate on a representation that is less sensitive to modest changes in data quality. 

\subsection{Original Data preparation}  \label{sec:data_prep}

Our process requires spectrum data from three classes of astronomical objects: transient sources, galaxies, and stars.  In the SDSS dataset, spectra are primarily classified into two categories, namely stars and galaxies.  To enable the identification of TDEs within large spectroscopic datasets, it is therefore necessary to additionally incorporate spectra of transient sources.

We further restrict the wavelength coverage to the range of 4000–7000~\AA, which effectively imposes a redshift constraint, particularly at the red end. This requirement ensures that key spectral features characteristic of confirmed TDEs, such as the broad He~\textsc{ii} and broad H$\alpha$ emission lines, fall within the analyzed wavelength range as much as possible.

\subsubsection{Transients}  \label{sec:data_prep_1}

The spectral data of transient sources used in this work are obtained from the \texttt{WISeREP} \citep{Yaron2012} Next Generation SuperFit (NGSF) database \citep{NGSF}  \footnote{\href{https://www.wiserep.org/content/}{\url{https://www.wiserep.org/content/}}}. Rather than applying the NGSF framework directly, we utilize its underlying spectral library, which was originally developed for template-matching applications.

The numbers of transient sources and spectra included in this database are summarized in Table~\ref{tbl:transients}.

\begin{deluxetable}{lcc}[h]
\tablecaption{Statistics of Transient Classes in the NGSF Spectral Library}
\tablehead{
\colhead{Class} &
\colhead{\ Number of Sources\ } &
\colhead{\ Number of Spectra\ }
}
\startdata
TDE H                &    3 &   13 \\
TDE H+He             &    5 &   30 \\
TDE He               &    2 &    4 \\
Ca-Ia                &    3 &   18 \\
Ca-Ib                &    7 &   34 \\
FBOT                 &    1 &    4 \\
II                   &   18 &  180 \\
II-flash             &    7 &   43 \\
IIb                  &   11 &  118 \\
IIb-flash            &    1 &    4 \\
IIn                  &    8 &   74 \\
ILRT                 &    4 &   21 \\
Ia 02es-like         &    4 &   26 \\
Ia 91T-like          &    4 &   29 \\
Ia 91bg-like         &    5 &   53 \\
Ia 99aa-like         &    2 &   19 \\
Ia-02cx like         &    5 &   34 \\
Ia-CSM               &    6 &   37 \\
Ia-CSM-(ambiguous)   &    2 &   17 \\
Ia-norm              &   13 &  124 \\
Ia-pec               &    3 &   36 \\
Ia-rapid             &    2 &    9 \\
Ib                   &   10 &   85 \\
Ibn                  &    7 &   60 \\
Ic                   &   10 &   81 \\
Ic-BL                &    8 &   67 \\
Ic-pec               &    1 &    9 \\
SLSN-I               &   17 &  142 \\
SLSN-II              &    3 &   21 \\
SLSN-IIb             &    1 &    6 \\
SLSN-IIn             &    6 &   32 \\
SLSN-Ib              &    1 &   12 \\
SN - Impostor        &    3 &   15 \\
computed             &    1 &   10 \\
super-chandra        &    4 &   47 \\
\enddata
\tablecomments{
The table summarizes the number of unique sources and the corresponding number of spectra for each transient class in the NGSF database. We emphasize the TDE classification in bold.
}
\label{tbl:transients}
\end{deluxetable}

\subsubsection{Galaxies}  \label{sec:data_prep_2}

The galaxy data used in this work are obtained from SDSS DR7. We perform uniform sampling based on the \texttt{SUBCLASS} field in the catalog. The SDSS galaxy \texttt{SUBCLASS} categories mainly include STARFORMING, STARBURST, AGN, AGN BROADLINE, and BROADLINE, as well as STARFORMING BROADLINE and STARBURST BROADLINE. 

In this work, we refer to samples with an empty \texttt{SUBCLASS} label as “Featureless”. In contrast, objects with a non empty \texttt{SUBCLASS} label generally show clear spectral line features.

We adopt a stratified sampling strategy and select 4000 spectra for each group. The motivation is explained as follows.

We employ the primary spectral of galaxy extracted by \citet{meansk86}, which provides an unsupervised taxonomy of 702,248 galaxy and quasar spectra ($z < 0.25$) from the SDSS DR7 dataset. This scheme organizes the spectral diversity into 86 principal types. For the purposes of this study, the mean spectra corresponding to these 86 classes are defined as our foundational templates, hereafter collectively referred to as the \texttt{meansk86} templates. 

Within this taxonomic structure, 27 classes are formally identified as star-forming galaxies, resulting in a maximum of 59 categories for the remaining populations. To guarantee that our sample maintains a statistically representative coverage across all potential sub-populations, we performed a sample size estimation. Under the conservative assumption of a symmetric (unbiased) Dirichlet distribution—wherein each of the 59 categories is assigned an equal prior probability—a minimum of 3,481 spectra is required to ensure that the expectation of selection for every category exceeds unity ($E[n_i] > 1$). To enhance the empirical robustness of our results and account for stochastic sampling effects, we adopted a final sample size of approximately 4,000 spectra per group.

Since our primary goal is the identification of TDEs, host galaxies exhibiting broad emission lines may introduce significant ambiguity after spectral superposition, because that TDEs intrinsically exhibit broad-line features, and therefore are more likely to be included in the non-AGN BROADLINE (including BROADLINE, STARFORMING BROADLINE, STARBURST BROADLINE) category.  If the selected BROADLINE samples are to contain genuine TDEs, labeling them as galaxies during the training and testing stages would introduce contamination, thereby affecting model performance and biasing the evaluation results. We therefore exclude galaxies labeled as BROADLINE from the superposition process which described in Section~\ref{sec:spectrum_mix}. In addition, given the increased complex properties of TDE in AGN hosts which revealed by both numerical simulation \citep{chan2019,chan2020} and observation \citep{PS16dtm,PS1-10adi,Liu2020,2019aalc,2022fpx}, galaxies classified as AGN are also excluded from the superposition step. Nevertheless, spectra classified as AGN and AGN BROADLINE are retained in the training set and model evaluation to ensure robust discrimination between TDEs and AGN-related transients. 


We emphasize that although the BROADLINE class has been excluded during model training and validation, these objects will still be included and classified in the final deployment of the model.

Given that our next analysis is restricted to the wavelength range of 4000–7000~\AA, we require all selected spectra to satisfy this coverage. 
We therefore first apply redshift corrections to the screened spectra. Only galaxy spectra that continue to fully cover the required wavelength range after correction are retained. 

At this stage, we note that after pre-screening, the number of \texttt{Featureless} sources decreases to 2,827, which is lower than the conservative estimate of 3,481 adopted previously. However, we consider that this difference does not significantly affect the results. According to \citealt{meansk86}, 27 classes are labeled as \texttt{Starforming}, while 12 classes of \texttt{Seyfert} are clearly identified as AGN. Therefore, the number of classes corresponding to \texttt{Featureless} should not exceed 47. Based on this estimate, only 2,209 sources are required to ensure sufficient coverage.

The final composition and distribution of the galaxy spectra used for training and testing are summarized in Table~\ref{tbl:Galaxy_sample}.

\subsubsection{Stars}  \label{sec:data_prep_3}

Although our analysis primarily focuses on galaxy samples from SDSS, a small fraction of stellar spectra may still be present in the dataset. When considering spectral information alone, the characteristic blue continuum of TDEs can be confused with that of early-type stars. We therefore explicitly include stellar spectra in our analysis to account for this potential source of contamination. The number of selected stellar samples is also summarized in Table~\ref{tbl:Galaxy_sample}.

\begin{deluxetable*}{lccc}[h]
\tablecaption{Statistics of SDSS Galaxies and Stars}
\tablehead{
\colhead{Class} &
\colhead{Number in Total} &
\colhead{Number in Our Sample} & 
\colhead{Number After Pre-screening}
}
\startdata
Featureless              &   618008 &  4000 & 2827 \\
STARFORMING               &    247684 &   4000 & 3604\\
STARBURST             &   65984 &  4000 & 3538 \\
AGN              &    16583 &   4000 & 3603 \\
BROADLINE              &   13334 &  0 & 0\\
AGN BROADLINE         & 2099 & 2099 & 1864\\
STARFORMING BROADLINE  &    1085 &   0 & 0 \\
STARBURST BROADLINE   &     111 &   0 & 0 \\
STAR & 488266 & 4000 & 3962
\enddata
\tablecomments{
Summary of the number of unique sources and the corresponding spectra for each galaxy \texttt{SUBCLASS} and for stars in the SDSS database and in our selected training sample. Pre-screening procedure based on rest-frame wavelength coverage, in which only spectra covering 4000–7000~\AA\ are retained.
}
\label{tbl:Galaxy_sample}
\end{deluxetable*}

\subsubsection{Data Pre-processing} \label{sec:pre-processing}

For galaxy and stellar spectra, we apply redshift correction to transform all spectra into the rest frame, using redshift values directly obtained from the SDSS DR7 FITS files. 


For transient source spectra, redshift correction is performed using the redshift information provided in the corresponding \texttt{WISeREP} table files.

Finally, all spectra are truncated to a wavelength range of 4000–7000~\AA\ and normalized via min–max scaling. To ensure spectral consistency and uniform dimensionality, each spectrum is linearly interpolated to a fixed grid of 4,000 data points.

\subsection{Evaluation Dataset Preparation}
\label{sec:evaluation}

Before applying the model to real observational data, it is necessary to first evaluate its performance. To this end, we construct a dedicated dataset for model validation. Specifically, the full dataset is initially divided into training and test subsets (Section~\ref{split}). From the training set, we independently derive PCA templates for galaxies, stars, and transient sources (Section~\ref{sec:pca_extraction}).
Subsequently, galaxy spectra and transient-source spectra are combined within both the training and test subsets (Section~\ref{sec:spectrum_mix}). Finally, PCA decomposition is performed on both the training and test sets by projecting all mixed spectra onto the fixed basis templates derived exclusively from the training data. (Section~\ref{sec:pca_generation}). 


\subsubsection{Training-Testing Data Split} \label{split}

To rigorously evaluate the performance of our method, we first enforce a strict separation between the training and test sets. 
Specifically, the spectra of transient sources, galaxies, and stars are stratified by class and randomly sampled, with 70$\%$ of the data assigned to the training set and the remaining 30$\%$ reserved for testing. In particular, for transient sources we will ensure that different spectra from the same source do not appear simultaneously in the training set and the testing set. The training and test samples are stored in separate directory structures. 



\subsubsection{PCA Templates Extraction} \label{sec:pca_extraction}

Prior to spectral mixing, PCA templates for galaxies, transient sources, and stellar spectra are derived exclusively from the training set and stored for later use. 
We employ \texttt{sklearn.decomposition.PCA} to perform PCA on the spectra of galaxies, transients, and stars. The implementation allows us to specify the desired number of principal components. In addition, the explained variance ratio is directly provided, which facilitates the determination of an appropriate number of components for the analysis.

For galaxy spectra, we retain 86 principal components, consistent with the \texttt{meansk86} \citep{meansk86} model adopted in this work, which provides a standardized representation of galaxy spectral diversity.

We do not directly use \texttt{meansk86} during model evaluation, as it may introduce a risk of data leakage. Specifically, \texttt{meansk86} is derived from a PCA of the entire SDSS spectral dataset, which includes spectra from our testing set. Consequently, its use could inadvertently incorporate information from the testing data, leading to overly optimistic performance estimates.

\begin{figure*}
\centering
\includegraphics[width=0.6\textwidth]{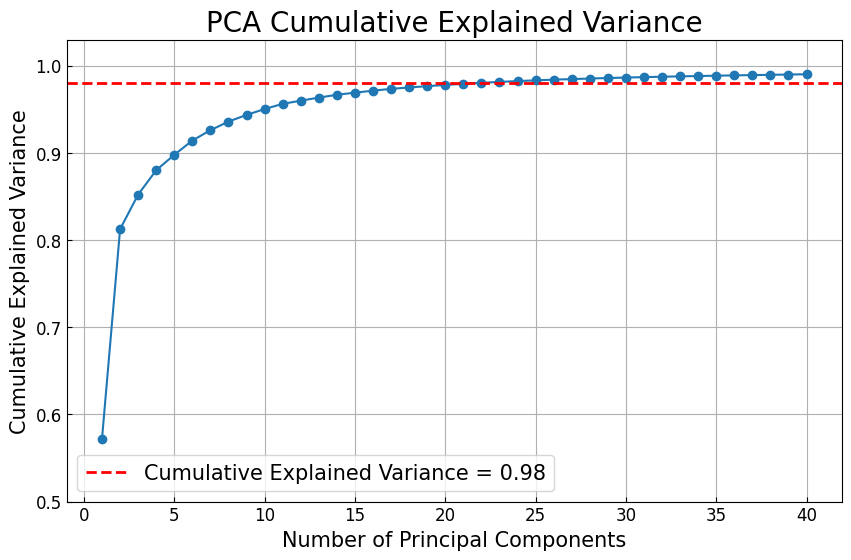}
\caption{Cumulative explained variance of the PCA for transient source spectra. Retaining 20 principal components captures approximately 98$\%$ of the total variance, indicating that the dominant spectral features are effectively preserved.
\label{fig:PCA_CEV}}
\end{figure*}

For stellar and transient source spectra, we retain 20 principal components, corresponding to approximately 98$\%$ of the cumulative explained variance (Figure~\ref{fig:PCA_CEV}). This choice achieves a balance between preserving the dominant spectral features and suppressing noise-dominated components, which is particularly advantageous for subsequent classification tasks. In this step, a total of 106 PCA components are derived from the training set.

\subsubsection{Transients-Galaxies Spectrum Mixing} \label{sec:spectrum_mix}

After constructing the datasets, we perform spectral superposition separately for the training and testing sets. Because transient phenomena including SNe and TDEs typically occur within host galaxies, contamination from host galaxy light is unavoidable in the observed spectra.  
To realistically represent this effect, we will construct composite spectra by superimposing transient source spectra onto host galaxy spectra with specified relative proportions. 

We consider the following classes of transient sources: TDEs (TDE-H, TDE-He, TDE-H+He), SN Ia (Ia-norm, Ia-91bg, Ia-91T, Ia-02es, Ia-02cx), SN Ib (Ib, Ibn), SN Ic (Ic), and SN II (II, IIn, IIb). For galaxy superposition, we restrict the sample to non-AGN and non-broad-line galaxies, including featureless, STARFORMING and STARBURST.

The core procedures involved in the preparation of the training and test datasets, namely spectral mixing and PCA decomposition are schematically illustrated in Figure~\ref{fig:processing_data}.

For each transient spectrum, we randomly select 100 host galaxy spectra from each host library (i.e., Featureless, STARFORMING, STARBURST). At the same time, a set of mixing scales is defined within a specified range (See Table~\ref{tbl:scale}). Each selected host spectrum is then paired with one unique scale value in a one-to-one manner, and this scale is drawn from a uniformly distributed random value within the given range. Synthetic mixed spectra are generated according to the following linear superposition in Eq~\ref{eq:flux_mixing}.

\begin{equation}
F_{\text{mix}}(\lambda) = F_{\text{transient}}(\lambda) + \mathrm{Scale} \cdot F_{\text{galaxy}}(\lambda)
\label{eq:flux_mixing}
\end{equation}

Where $F_{\text{mix}}(\lambda)$ represents the flux of the composite spectra, $F_{\text{transient}}(\lambda)$ denotes the rest-frame flux of the TDE (or other transient). $F_{\text{galaxy}}(\lambda)$ signifies the rest-frame flux of the host galaxy. $\mathrm{Scale}$ is the dimensionless mixing coefficient.

The relatively large number of certain types of transient sources may still lead to a substantial computational cost. After spectral superposition, we impose an upper truncation limit to reduce training cost. For each transient–galaxy pair, uniform random sampling is performed according to the scaling factor. The maximum number of superimposed spectra for each transient-source and galaxy combination is limited to 500.

The normalized galaxy and transient spectra are first superimposed according to the specified scale and then renormalized. 
To account for the altered noise properties introduced by post-superposition normalization, we incorporate supplementary Gaussian noise with a standard deviation of $0.2\sigma / (1 + \mathrm{Scale})$, where $\sigma$ denotes the standard deviation of the original spectrum and $\mathrm{Scale}$ represents the mixing ratio of the host galaxy components. This formulation compensates for the normalization-induced noise suppression by a factor of $(1 + \mathrm{Scale})^{-1}$ which would otherwise artificially inflate the SNR. The coefficient 0.2 is derived from a scaling constraint: in the limiting case where $\mathrm{Scale} = 0$, it ensures that a $1\sigma$ perturbation corresponds to a $5\sigma$ effective deviation following noise injection, thereby preserving the statistical significance of spectral fluctuations. This framework maintains a consistent noise interpretation across varying mixing ratios while preventing excessive amplification, effectively balancing noise restoration with the preservation of intrinsic spectral features.

After this step, wavelet-based denoising is applied to the noise-augmented mixed spectra using the \texttt{pywt} package \citep{Lee2019}. The Daubechies 4 (\texttt{db4}) wavelet family is employed, with a decomposition level of 2 and a threshold scale of 1.0. This parameter configuration effectively reduces noise while preserving the intrinsic spectral features. We emphasize that the noise injection in the previous step is not contradictory to the denoising applied here. The purpose of noise addition is to restore the noise level of the original spectra following normalization, whereas wavelet denoising aims to suppress noise components that may adversely affect the model performance. Although PCA-based dimensionality reduction already significantly mitigates the influence of noise, this step provides additional stability for high-order feature extraction. For later classification on the large-volume SDSS DR7 dataset, no artificial noise injection is applied.


The distribution of the mixed spectra in the training and test sets used for model performance evaluation is presented in Table~\ref{tbl:train_test_data}.

\begin{deluxetable*}{lccc}[ht]
\tablecaption{Mixing scale ranges adopted for different transient and galaxy combinations.\label{tab:mix_scale}}
\tablehead{
\colhead{Transient Type $\backslash$ Galaxy Type} &
\colhead{FEATURELESS} &
\colhead{STARFORMING} &
\colhead{STARBURST} 
}
\startdata
TDE      & 0.0--1.5 & 0.0--2.0 & 0.0--2.0  \\
SN Ia    & 0.0--2.0 & 0.0--10.0 & 0.0--10.0   \\
SN Ib   & 0.0--2.0 & 0.0--10.0 & 0.0--10.0   \\
SN Ic  & 0.0--2.0 & 0.0--10.0 & 0.0--10.0   \\
SN II    & 0.0--2.0 & 0.0--10.0 & 0.0--10.0   
\enddata
\tablecomments{
The mixing scale represents the relative flux contribution of the transient component
to the host galaxy spectrum within the wavelength range of 4000--7000~\AA.
}
\label{tbl:scale}
\end{deluxetable*}


\subsubsection{PCA Array Generation} \label{sec:pca_generation}

After constructing the superimposed spectra for both the training and test sets, we perform PCA on these spectra.  
For both the training and test sets, only the PCA templates derived from the training set are used to project the superimposed spectra, ensuring a strict separation between training and testing.  
We employ the \texttt{{lsq$\_$linear}} method from the \texttt{scipy} library, in conjunction with the PCA templates derived in Section~\ref{sec:pca_extraction}, to perform multi-dimensional linear fitting and extract the corresponding PCA coefficients.

To account for the superposition of host-galaxy and transient-source components, we construct a dual-channel PCA framework for spectral decomposition. The first channel consists of a joint basis set of 106 vectors, formed by concatenating the first 86 principal components (PCs) derived from galaxy spectra with the first 20 PCs obtained from transient sources. Complementary to this, the second channel utilizes 20 PCs extracted from a stellar training set to capture underlying stellar features.
Each superimposed spectrum is ultimately represented as a feature matrix with a shape of $(2, 106)$. The first row contains the 106 coefficients derived from the joint galaxy–transient basis, while the second row comprises the 20 stellar coefficients, with the remaining 86 entries zero-padded to maintain dimensional uniformity. This structured representation ensures a consistent input format for the subsequent classification model.


\begin{deluxetable}{lrrr}[h]
\tablecaption{The classification counts of each category in the training/testing set after spectral mixing}
\tablehead{
\colhead{Class} &
\colhead{Training Set} &
\colhead{Testing Set} &
\colhead{Total}
}
\startdata
TDE                &    1962 &   1356 & 3318 \\
SN Ia                &    3783 &   1974 & 5757 \\
SN Ib                &    1864 &   1316 & 3180 \\
SN Ic                 &    1120 &    611 & 1731 \\
SN II                   &   3140 &  2487 & 5627 \\
Galaxy  \  \   &   \    10805 &  \    4631 & \ \ \  15436\\
Star                   &   2775 &  1187 & 3962 \\
\enddata
\label{tbl:train_test_data}
\tablecomments{
Distribution of sample labels of mixed spectra in the training and testing sets used for model evaluation.
}
\end{deluxetable}

\subsection{Model Training and Evaluation}
\label{sec:Model Training and Evaluating}

Following the data preparation described in Section~\ref{sec:evaluation}, the resultant PCA-derived matrices are fed into the \texttt{MgFormer} architecture for model training and evaluation. \texttt{MgFormer} is a variant of the Transformer architecture, designed to effectively capture both inter-dimensional correlations and sequential dependencies within the input features \citep{Wen2024}. 

\texttt{Mgformer} is originally designed for Multivariate Time Series Classification (MTSC). By employing a multi-group architecture and a multi-head attention framework, \texttt{Mgformer} achieves specialized optimization for multivariate dimension classification, facilitating the joint modeling of multi-scale feature extraction and complex dimensional correlations. 
In our adapted implementation, the process is structured as follows:

Feature Embedding and Patching: After the input sequence is vectorized via the linear layers of $\texttt{trunk\_net}$, the resulting matrix is patched and transposed. This operation allows the model to capture multi-scale relationships both within individual sequences and across different sequence dimensions.

$\texttt{time-wise}$ Module: The core architecture comprises two primary components. The first is the $\texttt{time-wise}$ module, which incorporates positional encoding into the sequence after the initial linear mapping. This ensures that the structural and sequential information, essential for transformer-based architectures, is preserved.

$\texttt{channel-wise}$ and $\texttt{multi\_group\_encoders}$ Module: The second component is the $\texttt{channel-wise}$ module. Before entering this module, the latent vectors are segmented by the $\texttt{multi\_group\_encoders}$ module according to predefined hyperparameters. These segments are then transposed, allowing the channel-wise Module to extract inter-dimensional features.

Information Fusion: The outputs from the channel-wise module are integrated with the original $\text{multi\_group\_encoders}$ results via a residual-like summation. This mechanism ensures that the model simultaneously retains both intra-dimensional and inter-dimensional correlations.

Classification Head: Finally, the generated embedding vectors are passed to a fully connected multilayer perceptron (MLP) head for downstream classification.

This model has previously demonstrated strong performance in 
optical light-curve classification tasks 
\texttt{TTC} \citep{Zheng2025}. Although the present study does not directly address MTSC tasks, the inherent cross-dimensional representation learning capability of \texttt{Mgformer} remains highly effective for characterizing the complex dependencies within our feature space.

The hyperparameter settings adopted in this work are summarized in Table~\ref{tbl:hyperparameter}. 
Within the hyperparameter configuration, the multi$\_$group parameter is pivotal as it defines the number of segments into which \texttt{Mgformer} partitions the input data for feature extraction. Since our combined PCA vectors exhibit a dual part structure in both dimensions, we evaluate multi$\_$group values of $[1, 2]$, where 1 represents holistic extraction and 2 corresponds to a bifurcated segmental analysis. Other optimized parameters influencing convergence and performance include the batch size (batch), learning rate (lr), and the number of Transformer encoder layers and attention heads (n$\_$layers, nhead). Additional architectural specifications involve the hidden space dimensions (emb$\_$size, nhid), the masking ratio (masking$\_$ratio), and the high-attention interval selection (ratio$\_$highest$\_$attention).

Specifically, we adopt the default hyperparameter configuration established by \citet{Wen2024}, which has been optimized through a controlled-variable framework to ensure superior convergence across diverse datasets.


We employ the \texttt{Adam} optimizer \citep{Adam} and the cross-entropy loss for model training. Model performance is evaluated after each epoch using two metrics: the F1 score for TDE classification on the test set ($\mathrm{F1_{TDE}}$) and the overall F1 score across all classes ($\mathrm{F1_{total}}$). 
The model checkpoint corresponding to the epoch that maximizes $\mathrm{F1_{TDE}} + \mathrm{F1_{total}}$ is selected as the final model. 
This model is deployed for training and testing on NVIDIA A100-PCIE-40GB. Each training epoch takes about 15 minutes.

\begin{figure*}[htbp] 
\centering
\includegraphics[width=0.405\textwidth]{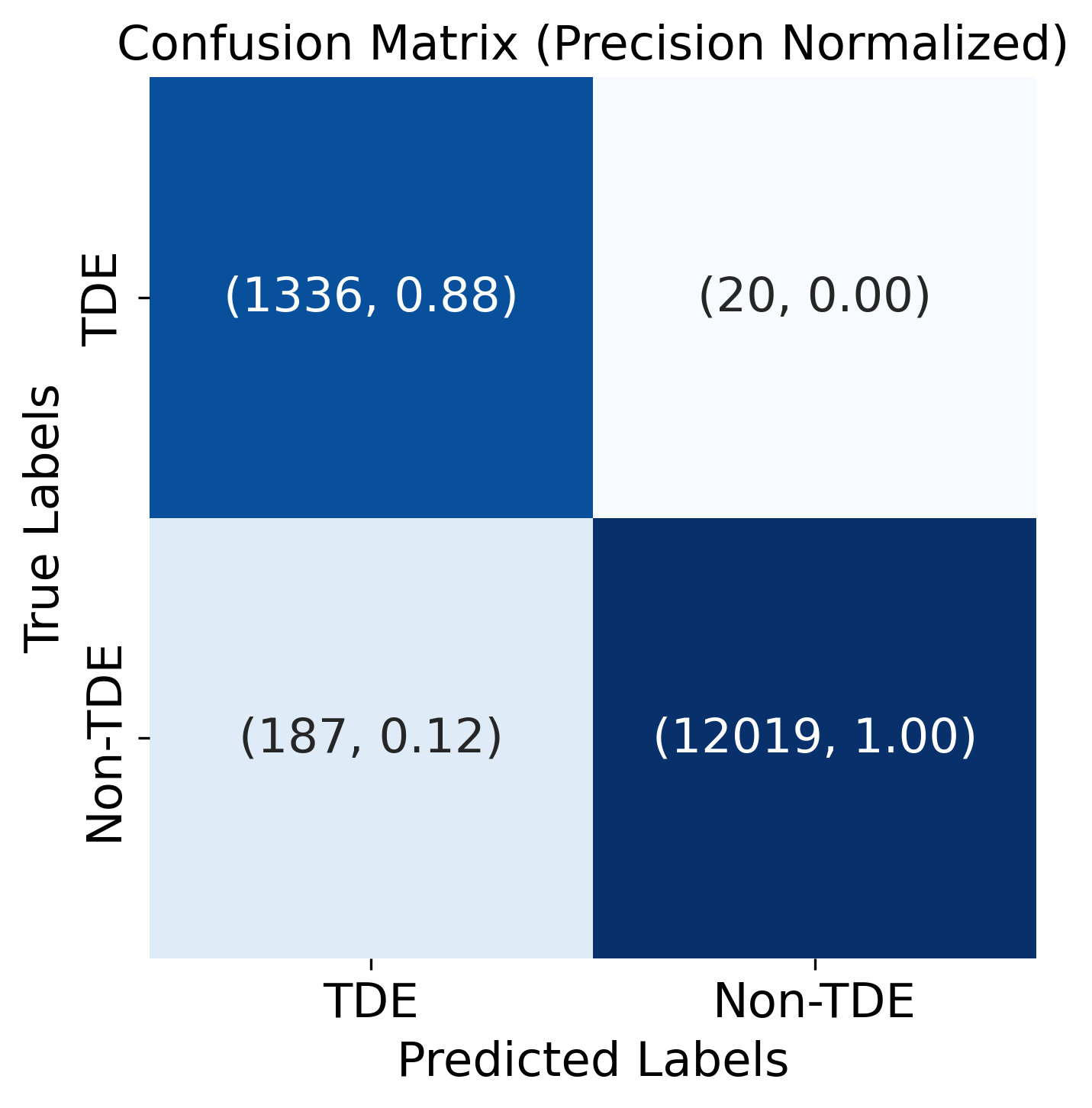}
\hspace{2cm}
\includegraphics[width=0.4\textwidth]{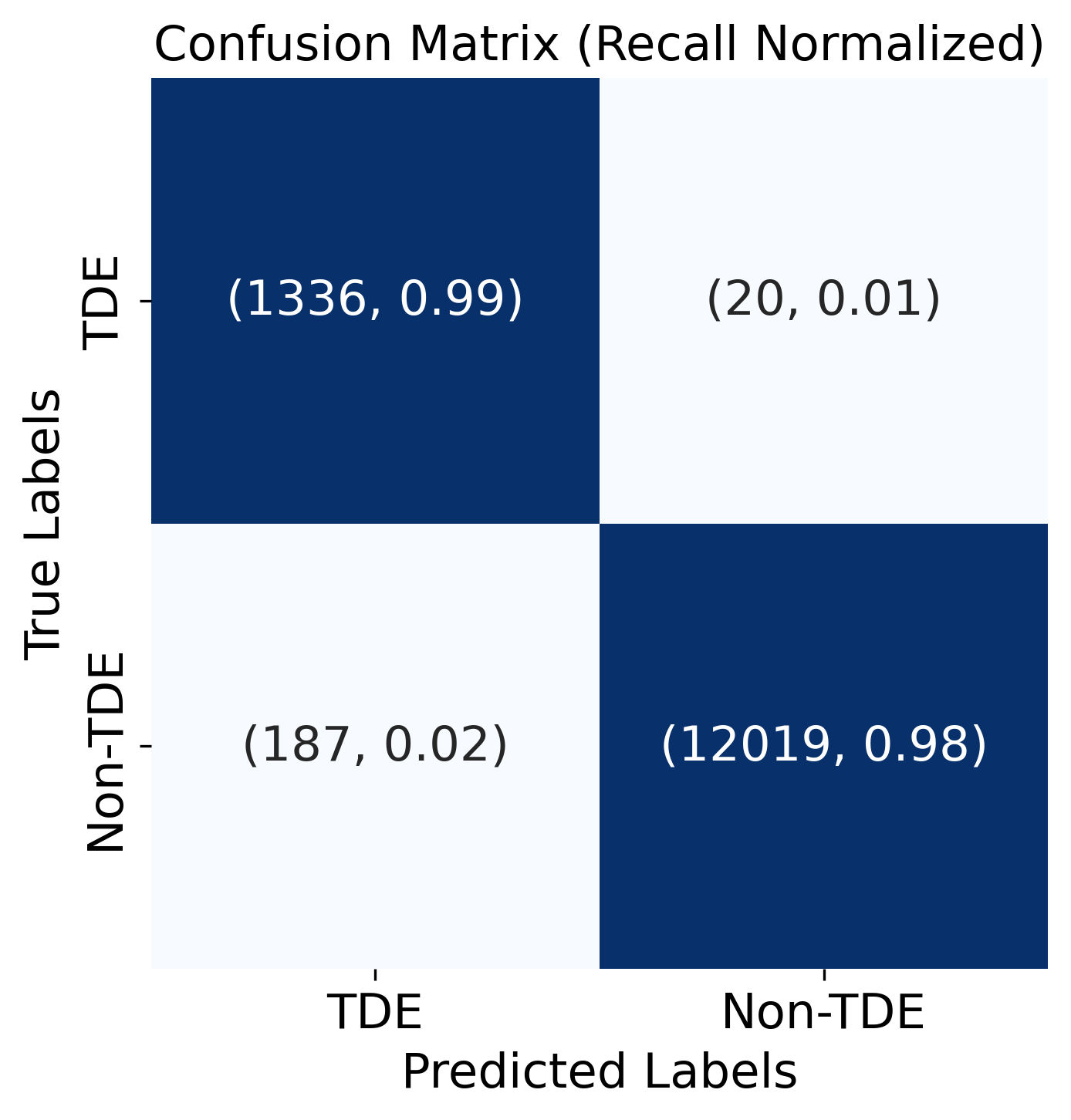}
\caption{The confusion matrices that show the capability of distinguishing TDEs from non-TDEs. For the independent test set used for performance evaluation, the validation model achieved a precision of 0.88 and a recall of 0.99 for TDE classification.}
\label{fig:2x2}
\end{figure*}

The results of the evaluation training are presented in Figure~\ref{fig:2x2}. On the test set, the model achieves a precision of 0.88 and a recall of 0.99.

The output classification vector quantifies the model's confidence across all target classes. We specifically utilize the scalar value for the TDE class, designated as the "TDE score", as the primary metric for candidate identification.

\begin{deluxetable}{c|c}[ht]
\centering
\tablecaption{Hyperparameters set for training}
\tablehead{
\colhead{\ \ Hyperparameters\ \ } &  \colhead{\ \ Value\ \ }
}
\startdata
multi$\_$group & [1,2]\\
batch & 8\\
lr &\ \ 0.0001\ \ \\
nlayers & 2\\
emb$\_$size &128\\
nhead & 8\\
emb$\_$size$\_$c & 128\\
masking$\_$ratio & 0.15\\
\ \ \ \ \ ratio$\_$highest$\_$attention\ \ \ \ \ \ & 0.35\\
dropout& 0.01\\
nhid& 128\\
nhid$\_$c& 128\\
Training epochs & 40
\enddata
\tablecomments{The definition and default setting of these hyperparameters are introduced in \citet{Wen2024}.}
\label{tbl:hyperparameter}
\end{deluxetable}

\subsection{Model Application}
\label{sec:real_model_tt}

\subsubsection{Model Re-training for Using in Real Data} \label{sec:real_model}

After completing the model evaluation, we retrain the model using the entire dataset to maximize the utility of the available spectra. At this stage, the partition between the training and test sets is eliminated, and the procedures for spectral superposition and PCA matrix generation are applied to the full dataset within a consolidated pipeline. To enhance the model's representational power during this final training, the PCA templates for host-galaxy decomposition are no longer restricted to those derived from the initial training set; instead, we adopt the more comprehensive \texttt{meansk86} library \citep{meansk86}.

\subsubsection{Classification in SDSS DR7 catalog} \label{sec:classification}


The classification tasks are conducted on a multi-core CPU system equipped with Intel Xeon Gold 6132 processors (4 sockets, 14 cores per socket, with 2 hardware threads per core, totaling 112 logical CPUs).  The model is deployed using a multi-process framework, where 20 processes are utilized for parallel classification.

Prior to classifying the full SDSS DR7 dataset, a series of pre-selection criteria are applied to ensure data quality and relevance. First, only sources with \texttt{OBJECTTYPE = GALAXY} are retained to minimize stellar contamination. Second, a median SNR greater than 15 is required to exclude low-quality spectra that could lead to unreliable classifications. Third, the rest-frame wavelength coverage is constrained such that the minimum wavelength is below 4000~\AA\ and the maximum wavelength exceeds 7000~\AA\ after redshift correction. In addition, to ensure a minimum level of spectral completeness, at least 1,000 data points are required within this interval.

Following the standardized preprocessing pipeline (min–max normalization, linear interpolation, and wavelet-based denoising), the spectra undergo PCA-based dimensionality reduction using the same templates established in Section~\ref{sec:real_model}. This ensures the input features remain consistent with the re-training stage of the classification model described therein.



\section{Results} \label{sec:result}

\subsection{From TDE candidates to reliable TDEs} \label{sec:candidates}

\begin{figure*}
\centering
\includegraphics[width=0.7\textwidth]{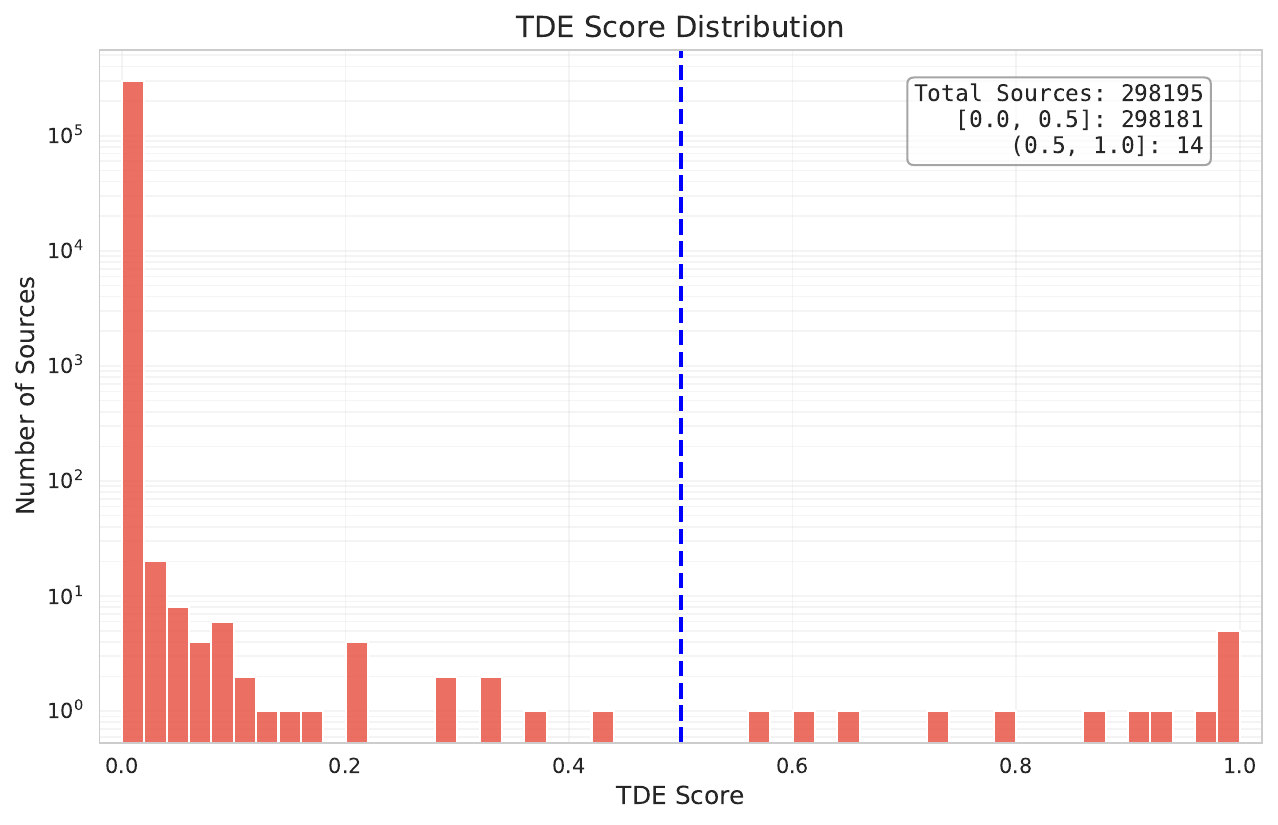}
\caption{TDE Score distribution across all evaluated sources.  The vast majority of the population is concentrated at low confidence levels ($<0.1$).  The blue dotted line indicates the selection threshold (TDE Score $= 0.5$), above which sources are identified as potential candidates.  Only 14 of the total samples have TDE score$>$0.5.
\label{fig:score_distribution}}
\end{figure*}

After applying selection cuts on the SDSS DR7 spectra, 298195 spectra are passed into the classifier. We adopt a threshold of TDE score $> 0.5$ for candidate selection. In a normalized classification space, a score exceeding 0.5 indicates that the confidence level for the TDE class surpasses the combined likelihood of all alternative classifications. The distribution of TDE Scores for the full sample is illustrated in Figure~\ref{fig:score_distribution}. Out of approximately 300,000 sources, only 14 candidates yielded a score greater than 0.5; their specific identifiers and corresponding scores are detailed in Table~\ref{tab:tde_candidate_score_ls}. Below are the procedures of the classification of these 14 TDE candidates, and Table~\ref{tab:class} provides a brief summary of these procedures and final classification.

Finally, we selected two newly discovered and confirmed TDEs and one previously reported likely TDE \citep{Wang2012} in this work. Their spectra are illustrated in Figure~\ref{fig:J124225.39+642919.0/J152459.70+045423.1-Fitting/J074820.67+471214.2}. Conversely, the spectra of the remaining 11 sources classified as non-TDEs or deemed improbable candidates following secondary screening are presented in Figure~\ref{fig:bogus}. The following subsections describe the screening procedure in detail.


\begin{deluxetable}{cc} 
\tablecaption{Classification Scores of Potential TDE Candidates \label{tab:tde_candidate_score_ls}}
\tablecolumns{2}
\tablewidth{0pt}
\tablehead{
\colhead{SDSS Name} & \colhead{TDE Score}
}
\startdata
J074820.66+471214.2 & 0.9833 \\
J124225.39+642919.0 & 0.7899 \\
J152459.70+045423.1 & 0.9999 \\
\midrule 
J004241.91+150926.1 & 0.7212 \\
J075525.29+391109.9 & 0.5787 \\
J094517.83+121327.9 & 0.9957 \\
J112110.21+530951.5 & 0.9828 \\
J113223.42+641958.4 & 0.6593 \\
J130728.92+563116.8 & 0.9096 \\
J140316.47+541649.9 & 0.9810 \\
J143751.21+064957.4 & 0.9306 \\
J152248.12+083148.0 & 0.6086 \\
J152851.89+000408.4 & 0.8771 \\
J214901.93-083026.6 & 0.9758 
\enddata

\tablecomments{SDSS names and TDE Scores for the 14 candidates exceeding the 0.5 threshold. The first three entries represent confirmed or highly probable TDEs identified in this work, followed by sources subsequently classified as unlikely TDEs, AGNs, or local star-forming regions. Detailed classification analysis for each source is provided in Table~\ref{tab:class}.}
\end{deluxetable}

\subsubsection{Emission line diagnostics} \label{sec:emission}

All redshifts in the catalog are confirmed to be correct. Two sources have extremely low redshifts ($z<0.005$), we exclude them after inspecting the SDSS images and confirm they were taken on the spirals of nearby galaxies. The blue continuum ($f_{\lambda}\propto\lambda^\alpha$ and $\alpha<0$) that resembles TDEs is probably the reason for the mis-classification, but it may be caused by the star formation activity in these two cases. Those sources showing blue continuum and broad Balmer, helium or Bowen fluorescence emission lines are favored, because they are in accordance with the optical TDEs. However, Seyfert 1 galaxies can also produce broad H and He emission lines and blue continua. Therefore, only the existence of Bowen fluorescence lines (e.g., N~\textsc{iii} $\lambda\lambda4100,4640$) will be considered as a decisive evidence. Strong and durable activities in AGNs or intense star formation activities can create strong narrow emission lines in the spectra, particularly [O~\textsc{iii}] $\lambda$5007, while TDEs need decades to create such strong emission lines. Therefore, strong [O~\textsc{iii}] $\lambda$5007 emission lines favor either AGNs or star-forming (SF) galaxies. 

\subsubsection{Burst \& stochastic variability diagnostics} \label{sec:burststo}

Optical-UV TDEs generate additional luminosity on the basis of the galaxy background during its flaring state. Therefore, if the spectra were taken during this state, the optical and UV luminosity would roughly rise and drop monotonically around this time. Instead, if stochastic variability persists over the full light curve, it favors the existence of AGNs. The stochastic variability can often be clearly seen in the mid-infrared (MIR) bands.
There are several cases in which a TDE also create a burst in MIR bands, which is called an MIR echo as it often lags behind the optical-UV outburst \citep[e.g.,][]{Jiang2016,vanVelzen2016,Dou2016}. In summary, targets showing persistent stochastic variability in any of UV, optical or MIR light curves are considered as AGNs, while targets showing burst-like features only around the observation of the SDSS spectrum are strongly favored. The inspection method of bursts and stochastic variability on all 12 candidates with $z > 0.005$ is demonstrated as follows. 


The observation times of these spectra are from MJD = 51672 to 54561, which were earlier than all optical time-domain surveys except for the Catalina Real-time Transient Survey (CRTS, \citealt{Drake2009}), which partially covered this time range. During this time range, GALEX visited the full sky for 1-2 times, and a small portion was revisited for several more times, which can capture a UV outburst by chance. Although years after the spectrum observation dates, the high-cadence light curves of the Zwicky Transient Facility (ZTF) optical survey can benefit the judgment of stochastic variability \citep{Bellm2019}.

For GALEX data, we use \texttt{gPhoton2}~\citep{Stclair2022} to download the GALEX cutouts, visually inspect the cutouts to define the source and background regions, and generate background-subtracted FUV- and NUV-band light curves. We perform the Galactic extinction correction by using the extinction law of \citet{Fitzpatrick1999}, the standard extinction curve with $R_V=A_V/ E(B-V)=3.1$ \citep{Osterbrock2006} and adopting the $E(B-V)$ values from~\citet{Planck2016}. 
For CRTS data, we retrieve the V-band light curve for each source from the CRTS catalog, correct the Galactic extinction and combine each light curve into 10-day bins. For ZTF data, we retrieve the g-, r- and i-band light curve for each source from the ZTF DR24 catalog, correct the Galactic extinction and combine each light curve into 3-day bins. Finally, we retrieve the WISE W1- (3.4 $\mu$m) and W2- (4.6 $\mu$m) band light curves and combine each light curve into $\sim$0.5-yr bins \citep{Wright2010}. All light curves are presented in Figure~\ref{fig:uv_opt_ir_lc}.

A burst is determined if monotonic rise trend presents within 0.5 yr before the spectrum observation date or monotonic fall trend within 2.0 yr after the spectrum observation date. In this definition, one source has a burst in both GALEX bands (SDSS~J124225.39+642919.0), one source has a burst in CRTS V band (SDSS~J152459.70+045423.1), and two sources have bursts in both WISE bands (SDSS~J074820.66+471214.2 and SDSS~J075525.29+391109.8).  

The stochastic variability diagnostics are performed in two folds. For the well-sampled optical band (CRTS-V, ZTF-g/r/i) light curves, Bayesian damped random walk (DRW) fitting is employed \citep{Kelly2009, MacLeod2010}. For the sparsely sampled MIR light curves, reduced $\chi^2$ is used to evaluate. 

For each optical light curve, we model the time variability as a DRW with a characteristic timescale $\tau_{\rm DRW}$, and a long-term standard deviation of variability $\sigma_{\rm DRW}$: For two epochs $t_i$ and  $t_i+\Delta t$, cov($t_i, t_i+\Delta t)\equiv S(\Delta t)= \sigma_{\rm DRW}^2\,$exp($-\Delta t/\tau_{\rm DRW}$); and run a Markov chain Monte Carlo (MCMC) fitting to get $\sigma_{\rm DRW}$ and $\tau_{\rm DRW}$. We use 32 walkers and set the chain length as 10000 steps, and discard the first 5000 steps to ensure that the chain has converged. The 3$\sigma$ lower limit (0.15-th percentile) for $\tau$, denoted as $\tau_{\rm DRW,3\sigma-low}$, is used for the judgment. For CRTS-V band, only epochs that are either more than 0.5 years earlier or more than 2.0 years later than the spectrum observation date are used in the DRW fitting. the light curve is binned by 10 days, so only sources with $\tau_{\rm DRW,3\sigma-low}>10$ days will be classified as AGNs. For ZTF g, r and i bands, the highest value is used for classification. The light curves are binned by 3 days, so only sources with $\tau_{\rm DRW,3\sigma-low}>3$ days will be classified as AGNs. 

For each MIR light curve, reduced $\chi^2$ ($\chi^2$/d.o.f.) is calculated. The larger $\chi^2$/d.o.f. between W1 and W2 bands will be adopted for evaluation. Sources that have $\chi^2$/d.o.f. $>10$ will be classified as AGNs. In addition, we also look into the W1$-$W2 color, and classify those with W1$-$W2 $>$ 0.8 as AGNs \citep{Stern2012}. 

The above procedures yield two reliable TDEs: SDSS J124225.39+642919.0 and SDSS~J152459.70+045423.1, one likely TDE SDSS~J074820.66+471214.2, six AGNs, and three sources that are unlikely to be TDEs (Table~\ref{tab:class}). We look up the literature and find that both reliable TDEs have not been reported, but the likely TDE was discovered and reported by \citet{Wang2012}. 
The SDSS spectra of three TDEs/likely TDEs are shown in Figure~\ref{fig:J124225.39+642919.0/J152459.70+045423.1-Fitting/J074820.67+471214.2}.
The details of two newly discovered TDEs are introduced below.

\begin{table*}[htb!]
\footnotesize
\centering
\begin{threeparttable}\caption{Classification of 14 Model Selected TDE Candidates\label{tab:class}}
\doublerulesep 0.1pt \tabcolsep 3.7pt 
\begin{tabular*}{\textwidth}{ccccccccccc}
\hline
SDSS Name & $z$ & Blue & Broad & Strong & Burst & $\tau_{\rm DRW,3\sigma-low}$ (d) & $\tau_{\rm DRW,3\sigma-low}$ (d) & $\chi^2$/d.o.f. & W1$-$W2 & Class\\
 &  & Cont. & Lines & [O~\textsc{iii}] & During & CRTS-V & max(ZTF-g/r/i) & max(W1/W2) & WISE &  \\
\hline
J004241.90+150926.1 & 0.10146 & Yes & H & Yes & No & 11.7 & 6.0 & 22.6 & 0.53 & AGN\\
J074820.66+471214.2 & 0.06153 & ? & H?+He~\textsc{ii} & No & WISE & 0.1 & 0.1 & 453.2 & 0.25 & Likely TDE\\
J075525.29+391109.8 & 0.03354 & Yes & H & Yes & WISE & 37.1 & 43.0 & 160.2 & 0.63 & AGN\\
J094517.82+121327.9 & 0.13299 & No & No & No & No & 0.1 & 1.4 & 1.7 & 0.18 & Unlikely TDE\\
J112110.21+530951.5 & 0.00362 & Yes & No & Yes & - & - & - & - & - & Local SF Region\\
J113223.43+641958.4 & 0.20978 & Yes & H/He~\textsc{ii} & Yes  & No & 138.1 & 76.3 & 376.9 & 0.86 & AGN\\
J124225.39+642919.0 & 0.04246 & Yes & No & No & GALEX & 0.1 & 0.1 & 3.1 & 0.03 & TDE\\
J130728.91+563116.8 & 0.10382 & ? & H & Yes & No & 76.1 & 49.2 & 145.0 & 0.62 & AGN\\
J140316.47+541649.9 & 0.00060 & Yes & No & Yes & - & - & - & - & - & Local SF Region\\
J143751.20+064957.3 & 0.08578 & Yes & H & Yes & No & 20.8 & 6.0 & 65.7 & 0.52 & AGN\\
J152248.12+083148.0 & 0.03661 & Yes & No & No & No & 0.1 & 0.1 & 1.5 & 0.1 & Unlikely TDE\\
J152459.70+045423.1 & 0.04154 & Yes & H+He+Bowen & No & CRTS & 0.1 & 1.2 & 3.7 & 0.07 & TDE\\
J152851.88+000408.5 & 0.08972 & Yes & No & Yes & No & 0.1 & 0.1 & 2.0 & 0.13 & Unlikely TDE\\
J214901.92-083026.6 & 0.17248 & No & H$\alpha$ & Yes & No & 0.8 & 2.4 & 136.2 & 0.91 & AGN\\
\hline
\end{tabular*}
\begin{tablenotes}
\item 1. $z$: We inspect the images of two sources that have particularly low $z$, J112110.21+530951.5 and J140316.47+541649.9, and find that the spectra were taken on the spiral of nearby galaxies. Given their blue continua, we conclude that they are local star-forming (SF) regions.
\item 2. Blue Cont.: If the blue continuum exists, i.e., $f_{\lambda}\propto\lambda^\alpha$ and $\alpha<0$. Those objects that have $\alpha\sim0$ are marked in ``?". A blue continuum can be found in TDEs, AGNs and SF galaxies, which may lead to the mis-classification of the classifier.
\item 3. Broad Lines: The broad emission lines in the spectra. Only the existence of Bowen emission lines is regarded as a decisive evidence of TDEs. 
\item 4. Strong [O~\textsc{iii}]: Strong [O~\textsc{iii}] $\lambda$5007 emission lines often imply strong and durable activities in the cores of galaxies, which are typically induced by AGNs. Although TDEs can occur in AGNs, this evidence favors an impostor AGN. Nonetheless, narrow [O~\textsc{iii}] emission can also be produced in SF galaxies, so this evidence is not decisive for an AGN classification.
\item 5. Burst During: If a burst like feature presented around the date that SDSS spectrum was obtained (-0.5 to +2.0 yrs) in any of optical (CRTS), UV (GALEX) and MIR (WISE) bands.
\item 6. $\tau_{\rm DRW,3\sigma-low}$: the Bayesian 3$\sigma$ lower limit of the characteristic timescale of the damped random walk (DRW) given by Markov chain Monte Carlo (MCMC) fitting. For CRTS-V band, only epochs that are either more than 0.5 years earlier or more than 2.0 years later than the spectrum observation date are used in the DRW fitting. the light curve is binned by 10 days, so only $\tau_{\rm DRW,3\sigma-low}>10$ days will be classified as AGNs. For ZTF g, r and i bands, the highest value is used for classification. The light curves are binned by 3 days, so only $\tau_{\rm DRW,3\sigma-low}>3$ days will be classified as AGNs. 0.1~d is set as the lower boundary.
\item 7. $\chi^2$/d.o.f. (WISE): For WISE W1 and W2 bands, the larger $\chi^2$/d.o.f. value will be accepted. Sources that have $\chi^2$/d.o.f. $>$ 10 will be classified as AGNs.
\item 8. W1$-$W2: The difference between the Vega magnitudes of W1 and W2 bands. Sources that have W1$-$W2 $>0.8$ will be classified as AGNs.
\end{tablenotes}
\end{threeparttable}
\end{table*}

\begin{figure*}[ht]
\centering
\includegraphics[width=0.4\textwidth]{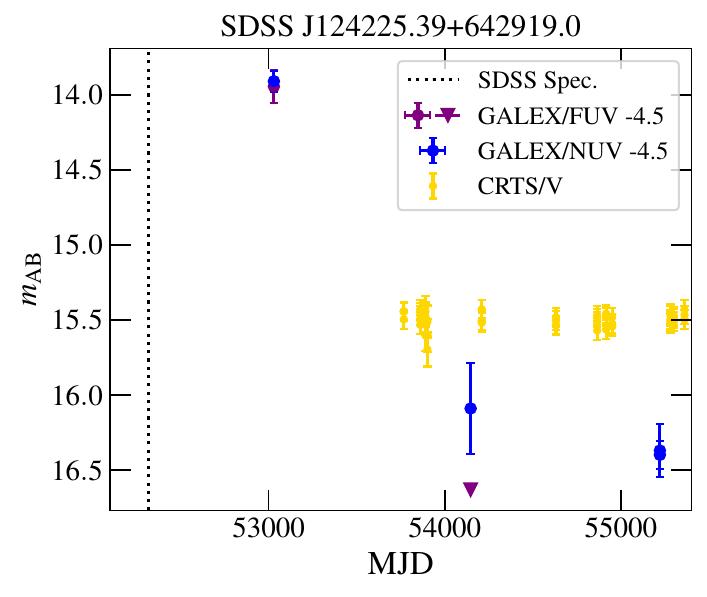}
\hspace{1cm}
\includegraphics[width=0.42\textwidth]{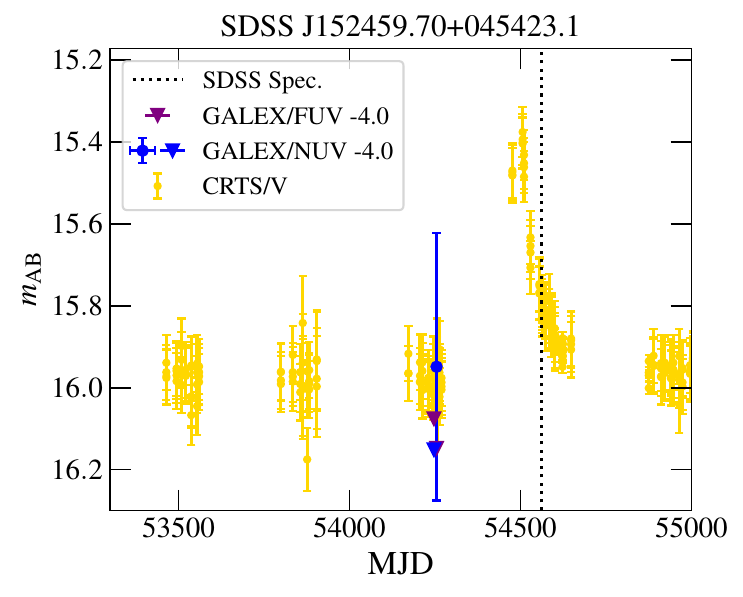}\\
\vspace{0.2cm}
\includegraphics[width=0.65\textwidth]{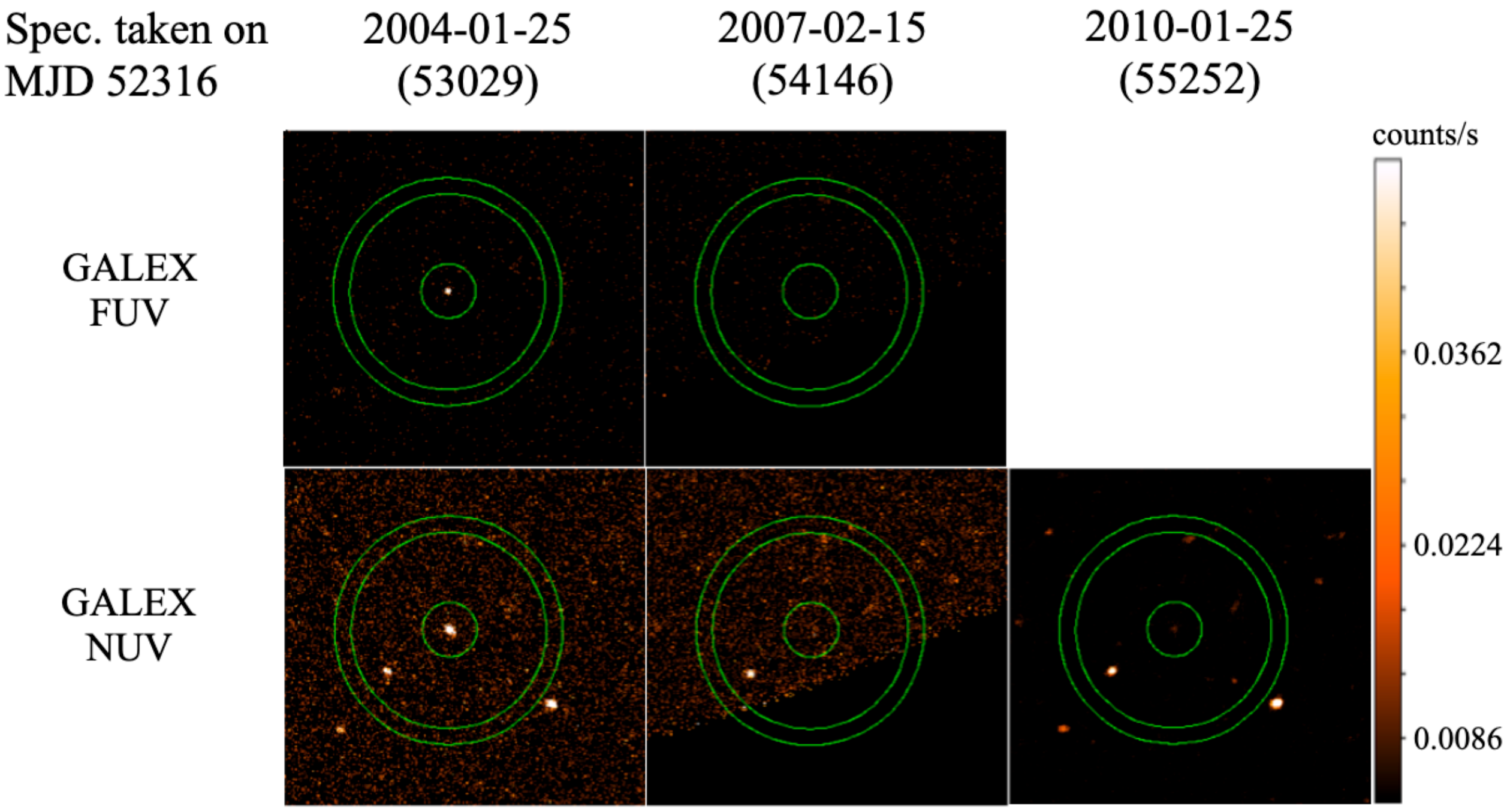}
\caption{\textit{Top: }UV (GALEX) and optical (CRTS) light curves of two newly discovered TDEs around the SDSS spectrum observation time (dotted black line). Magnitudes are corrected for Galactic extinction but not host-subtracted, and on which offsets are applied for clearer display. \textit{Bottom:} The GALEX FUV and NUV ``intensity'' map of SDSS~J124225.39+642919 in three epochs, which is created by dividing the count map by the relative response map. The green circle and annulus represent the source and background region that used in the light curve creation, respectively. They are centered at the position of SDSS~J124225.39+642919. The radius for the source region is 30$^{\prime\prime}$, and the inner and outer radius of the background region are 108$^{\prime\prime}$ and 126$^{\prime\prime}$, respectively. The map confirms that a transient UV source was present when the SDSS spectrum was taken.
\label{fig:lc_2tdes+map_tde1}} 
\end{figure*}

\subsection{New TDE 1: SDSS~J124225.39+642919.0}

We identify SDSS~J124225.39+642919.0 as a TDE based on evidence from both the spectrum and the light curve evolution. SDSS~J124225.39+642919.0 exhibits spectral features consistent with those expected for TDEs. These features include a blue continuum, broad Balmer emission lines, and the absence of strong narrow emission lines (e.g., [O~\textsc{iii}]~$\lambda$5007) which can exclude strong AGN activity~\citep{17TDE,distinguish_TDE}. The optical spectroscopic and UV features are self-consistent as the blue continuum can extend to the UV wavelength range. They can be naturally explained by a TDE scenario, while neither a supernova (SN) scenario nor a AGN scenario since the features have been never observed in these two types.

We determine that SDSS~J124225.39+642919.0 is the earliest optical–UV TDE discovery reported so far, primarily based on its light curve and the observation time of the SDSS spectrum.
The spectrum of SDSS~J124225.39+642919.0 taken on MJD = 52316 is dominated by a blue continuum. Its redshift can be tightly constrained by a number of absorption lines: $z=0.04240\pm0.00002$. The GALEX FUV and NUV map and light curve undoubtedly confirm that a UV source was present when the spectrum was taken, since both FUV- and NUV-band light curves reveal a bright source of 18.5 mag in the first epoch, and this source was probably caused by a transient, since it was no longer detected in FUV ($>$22 mag) and $\sim$2.5 mag fainter in the next epoch three years later (Top left panel of Figure~\ref{fig:lc_2tdes+map_tde1}). The presence and disappearance of this UV transient are clearly displayed in the GALEX cutouts (Bottom panel of Figure~\ref{fig:lc_2tdes+map_tde1}).  The TDE should start no later than the spectrum taken time, MJD = 52316. Hence, we conclude that this optical-UV outburst in SDSS~J124225.39+642919.0 should be the earliest optical-UV TDE discovered by now.

In the spectrum, weak H$\alpha$ and H$\delta$/N~III emission features can be recognized. The He~\textsc{ii}/H$\beta$ complex presents but can hardly be resolved. Nevertheless, we still attempt a two-component fit onto the He~\textsc{ii}/H$\beta$ complex, motivated by the potential bimodal structure in this wavelength range.

The FWHMs of each emission line are then directly derived from the best-fit Gaussian parameters. as shown in the top panel of Figure~\ref{fig:J124225.39+642919.0/J152459.70+045423.1-Fitting/J074820.67+471214.2}. The black solid line represents the observed spectrum, while the gray dashed line denotes the continuum fitted with a third-order polynomial. The purple and orange shaded regions correspond to the broad H$\alpha$ and He~\textsc{i} emission components, respectively. The light blue shaded region indicates the broad H$\beta$ + He~\textsc{ii} emission, and the green shaded region represents the H$\delta$/N~\textsc{iii} emission component. The red solid line shows the total best-fit model obtained by summing the continuum and all emission-line components.

The best-fit FWHMs for the H$\alpha$, H$\beta$, H$\delta$/N~\textsc{iii}~$\lambda$4100 and He~\textsc{ii}~$\lambda$4686 emission lines are 9663$\pm$934 km~s$^{-1}$, 16709$\pm$4105 km~s$^{-1}$, 11524$\pm$1753 km~s$^{-1}$ and 14276$\pm$2850 km~s$^{-1}$, respectively.


\subsection{New TDE 2: SDSS~J152459.70+045423.1}

The spectrum of SDSS~J152459.70+045423.1 was taken on MJD = 54562. Its redshift can be tightly constrained by the Ca \textsc{ii} doublets: $z=0.04154\pm0.00024$ \footnote{In data releases later than DR7, the source redshift is mistakenly fitted to $z=1.44558$.}, and proven by the presence of H$\alpha$, H$\beta$, H$\delta$/N~\textsc{iii}~$\lambda$4100, He~\textsc{ii}~$\lambda$4686 and He~\textsc{i}~$\lambda$5876 emission lines. 

In the spectrum of SDSS J152459.70+045423.1, we clearly identify broad emission lines of H$\alpha$, H$\beta$, H$\delta$/N~\textsc{iii}~$\lambda$4100, He~\textsc{ii}~$\lambda$4686, and He~\textsc{i}~$\lambda$5876. These broad emission lines and the blue continuum strikingly resemble a typical TDE-H+He spectrum. In the CRTS V-band light curve, we find there was a clear outburst when the spectrum was taken (Top right panel of Figure \ref{fig:lc_2tdes+map_tde1}). Hence, we conclude that this optical outburst in SDSS~J152459.70+045423.1 is most likely to be a TDE-H+He.



\begin{figure*}[h]
\centering

\includegraphics[width=0.805\textwidth]{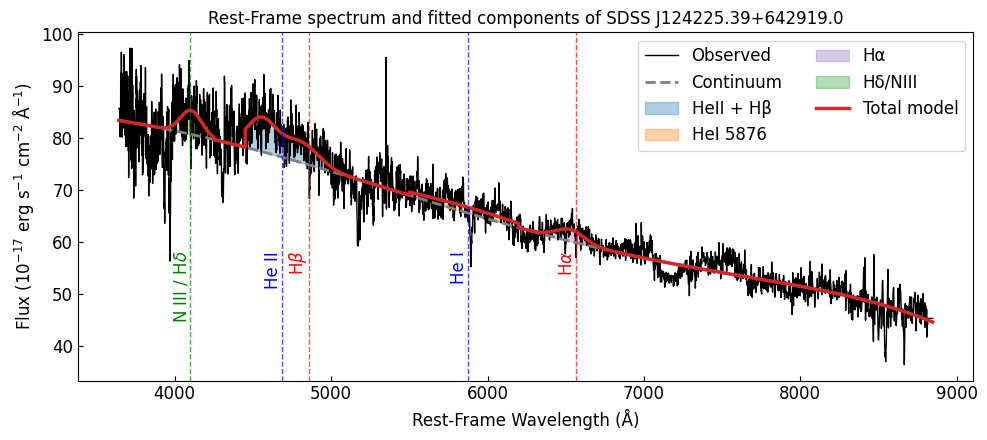}

\vspace{0.5cm}

\includegraphics[width=0.8\textwidth]{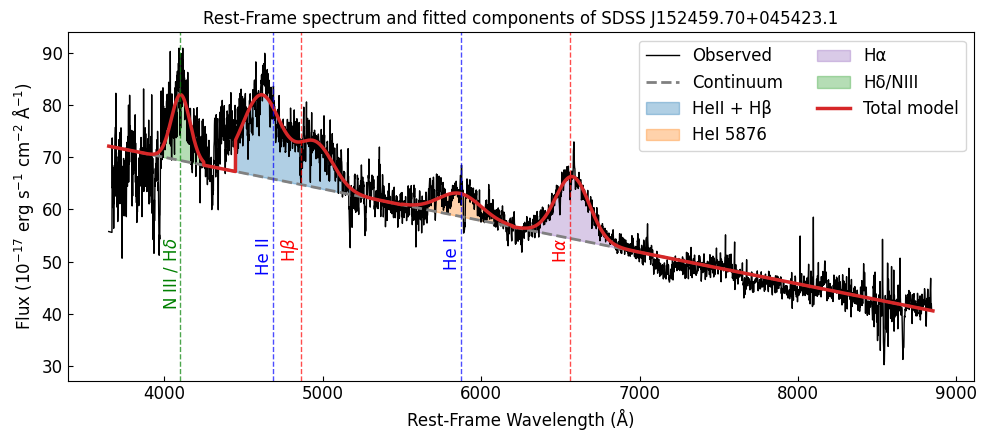}

\vspace{0.5cm}

\includegraphics[width=0.8\textwidth]{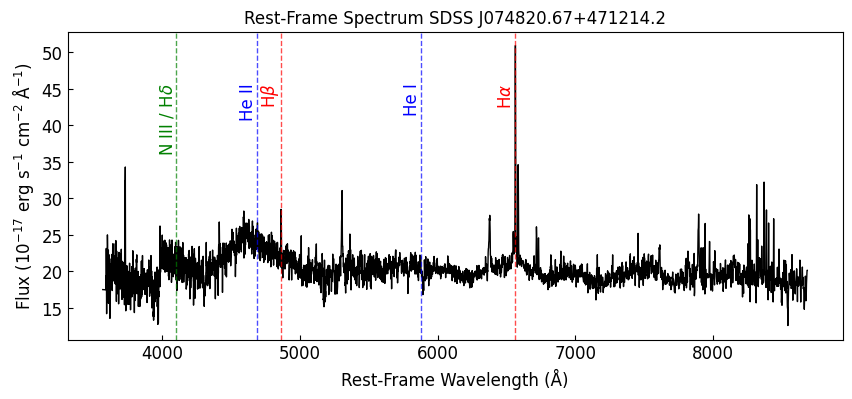}

\caption{The rest-frame spectra of SDSS J124225.39+642919.0, SDSS J152459.70+045423.1 and SDSS J074820.67+471214.2. In particular, SDSS J152459.70+045423.1 and SDSS J124225.39+642919.0 are first reported in this article and spectral fitting is performed on them. The best-fit continuum is plotted as a gray dashed line. Shaded regions indicate individual emission-line components: broad H$\alpha$ (purple) and He~\textsc{i} (orange), broad H$\beta$ + He~\textsc{ii} (light blue), and H$\delta$/N~\textsc{iii} (green). The red solid line shows the total model combining the continuum and all emission lines.}
\label{fig:J124225.39+642919.0/J152459.70+045423.1-Fitting/J074820.67+471214.2}
\end{figure*}

Similar to SDSS J124225.39+642919.0, the spectrum and spectral-line fitting of SDSS J152459.70+045423.1 are shown in the middle panel of Figure~\ref{fig:J124225.39+642919.0/J152459.70+045423.1-Fitting/J074820.67+471214.2}.

The best-fit FWHMs for the H$\alpha$, H$\beta$, H$\delta$/N~\textsc{iii}~$\lambda$4100, He~\textsc{ii}~$\lambda$4686 and He~\textsc{i}~$\lambda$5876 emission lines are 11444$\pm$172 km~s$^{-1}$, 16128$\pm$1028 km~s$^{-1}$, 10133$\pm$588 km~s$^{-1}$, 18440$\pm$809 km~s$^{-1}$ and 16193$\pm$813 km~s$^{-1}$, respectively. 




\section{Summary and Future Application} \label{sec:conclusion}

In this work, we build a Transformer-based TDE spectrum classifier, and this classifier successfully digs out two reliable TDEs in the SDSS DR7 catalog that have not been discovered by previous works and one likely TDE that have been reported. The combination of TDE-like spectrum and contemporary optical or UV counterparts strongly confirms the TDE nature of these two newly discovered TDEs. Particularly, the TDE flare in SDSS~J124225.39+642919.0 occurred earlier than the spectrum observation time, MJD = 52316 (February 11, 2002), making it the earliest optical-UV TDE discovered by now. The other TDE in SDSS~J152459.70+045423 occurred later but is still among the earliest optical-UV TDEs. 
The discovery of these two new TDEs in the widely used SDSS DR7 catalog highlights the power of machine learning based classifiers in digging out buried treasures in large-volume catalogs, and offers a new method for discovering optical-UV TDEs.

In the future works, we will explore the TDE candidates in larger spectroscopic catalogs, e.g., DESI DR1 catalog \citep{DESIDR1}. The redshift distribution will be towards higher redshifts, hence the emission line features that determine the transient types will be those in shorter wavelengths. We will add more TDE spectra with higher redshifts, and adjust the selection criteria of the galaxy spectra. Comparing with the time that SDSS spectra were taken, nowadays we can utilize much more abundant wide-field time-domain survey data to judge the TDE candidates. Regarding that we are currently standing on the surge of the big spectroscopic survey era, the prospect of the TDE spectrum classifier is promising and may bring exciting results in the next few years.

The classification pipeline developed for this study is open-source and can be accessed via GitHub at \url{https://github.com/Tico-Astro/PCA-Transformer-Spectrum-Classification-Pipeline.}

\begin{acknowledgments}
This work is supported by the National Science Foundation of China (NSFC, Grant No. 12233008, 12473008), the National Key R\&D Program of China (2023YFA1608100, 2025YFF0511003), the Strategic Priority Research Program of the Chinese Academy of Sciences (Grant No. XDB0550200), the Cyrus Chun Ying Tang Foundations, and the 111 Project for "Observational and Theoretical Research on Dark Matter and Dark Energy" (B23042). The CRTS survey is supported by the U.S.~National Science Foundation under grants AST-0909182. Z.L. is supported by the DIM ORIGINES funded by the Region Île-de-France (grant IDF-DIM-ORIGINES-2024-2-01).
\end{acknowledgments}


\vspace{5mm}

\software{\texttt{Astropy} \citep{astropy},  
          \texttt{Scipy} \citep{scipy},
          \texttt{Pytorch} \citep{pytorch2},
          \texttt{Matplotlib} \citep{matplotlib},
          \texttt{Pywt} \citep{Lee2019},
          }

\bibliography{sample701}{}

\begin{thebibliography}{}
\expandafter\ifx\csname natexlab\endcsname\relax\def\natexlab#1{#1}\fi
\providecommand{\url}[1]{\href{#1}{#1}}
\providecommand{\dodoi}[1]{doi:~\href{http://doi.org/#1}{\nolinkurl{#1}}}
\providecommand{\doeprint}[1]{\href{http://ascl.net/#1}{\nolinkurl{http://ascl.net/#1}}}
\providecommand{\doarXiv}[1]{\href{https://arxiv.org/abs/#1}{\nolinkurl{https://arxiv.org/abs/#1}}}

\bibitem[{K.~N. {Abazajian} {et~al.}(2009){Abazajian}, {Adelman-McCarthy},
  {Ag{\"u}eros}, {Allam}, {Allende Prieto}, {An}, {Anderson}, {Anderson},
  {Annis}, {Bahcall}, {Bailer-Jones}, {Barentine}, {Bassett}, {Becker},
  {Beers}, {Bell}, {Belokurov}, {Berlind}, {Berman}, {Bernardi}, {Bickerton},
  {Bizyaev}, {Blakeslee}, {Blanton}, {Bochanski}, {Boroski}, {Brewington},
  {Brinchmann}, {Brinkmann}, {Brunner}, {Budav{\'a}ri}, {Carey}, {Carliles},
  {Carr}, {Castander}, {Cinabro}, {Connolly}, {Csabai}, {Cunha}, {Czarapata},
  {Davenport}, {de Haas}, {Dilday}, {Doi}, {Eisenstein}, {Evans}, {Evans},
  {Fan}, {Friedman}, {Frieman}, {Fukugita}, {G{\"a}nsicke}, {Gates},
  {Gillespie}, {Gilmore}, {Gonzalez}, {Gonzalez}, {Grebel}, {Gunn},
  {Gy{\"o}ry}, {Hall}, {Harding}, {Harris}, {Harvanek}, {Hawley}, {Hayes},
  {Heckman}, {Hendry}, {Hennessy}, {Hindsley}, {Hoblitt}, {Hogan}, {Hogg},
  {Holtzman}, {Hyde}, {Ichikawa}, {Ichikawa}, {Im}, {Ivezi{\'c}}, {Jester},
  {Jiang}, {Johnson}, {Jorgensen}, {Juri{\'c}}, {Kent}, {Kessler}, {Kleinman},
  {Knapp}, {Konishi}, {Kron}, {Krzesinski}, {Kuropatkin}, {Lampeitl},
  {Lebedeva}, {Lee}, {Lee}, {French Leger}, {L{\'e}pine}, {Li}, {Lima}, {Lin},
  {Long}, {Loomis}, {Loveday}, {Lupton}, {Magnier}, {Malanushenko},
  {Malanushenko}, {Mandelbaum}, {Margon}, {Marriner}, {Mart{\'\i}nez-Delgado},
  {Matsubara}, {McGehee}, {McKay}, {Meiksin}, {Morrison}, {Mullally}, {Munn},
  {Murphy}, {Nash}, {Nebot}, {Neilsen}, {Newberg}, {Newman}, {Nichol},
  {Nicinski}, {Nieto-Santisteban}, {Nitta}, {Okamura}, {Oravetz}, {Ostriker},
  {Owen}, {Padmanabhan}, {Pan}, {Park}, {Pauls}, {Peoples}, {Percival}, {Pier},
  {Pope}, {Pourbaix}, {Price}, {Purger}, {Quinn}, {Raddick}, {Re Fiorentin},
  {Richards}, {Richmond}, {Riess}, {Rix}, {Rockosi}, {Sako}, {Schlegel},
  {Schneider}, {Scholz}, {Schreiber}, {Schwope}, {Seljak}, {Sesar}, {Sheldon},
  {Shimasaku}, {Sibley}, {Simmons}, {Sivarani}, {Allyn Smith}, {Smith},
  {Smol{\v{c}}i{\'c}}, {Snedden}, {Stebbins}, {Steinmetz}, {Stoughton},
  {Strauss}, {SubbaRao}, {Suto}, {Szalay}, {Szapudi}, {Szkody}, {Tanaka},
  {Tegmark}, {Teodoro}, {Thakar}, {Tremonti}, {Tucker}, {Uomoto}, {Vanden
  Berk}, {Vandenberg}, {Vidrih}, {Vogeley}, {Voges}, {Vogt}, {Wadadekar},
  {Watters}, {Weinberg}, {West}, {White}, {Wilhite}, {Wonders}, {Yanny}, \&
  {Yocum}}]{SDSSDR7}
{Abazajian}, K.~N., {Adelman-McCarthy}, J.~K., {Ag{\"u}eros}, M.~A., {et~al.}
  2009, \bibinfo{title}{{The Seventh Data Release of the Sloan Digital Sky
  Survey},} \apjs, 182, 543, \dodoi{10.1088/0067-0049/182/2/543}

\bibitem[{I. {Arcavi} {et~al.}(2014){Arcavi}, {Gal-Yam}, {Sullivan}, {Pan},
  {Cenko}, {Horesh}, {Ofek}, {De Cia}, {Yan}, {Yang}, {Howell}, {Tal},
  {Kulkarni}, {Tendulkar}, {Tang}, {Xu}, {Sternberg}, {Cohen}, {Bloom},
  {Nugent}, {Kasliwal}, {Perley}, {Quimby}, {Miller}, {Theissen}, \&
  {Laher}}]{Arcavi2014}
{Arcavi}, I., {Gal-Yam}, A., {Sullivan}, M., {et~al.} 2014, \bibinfo{title}{{A
  Continuum of H- to He-rich Tidal Disruption Candidates With a Preference for
  E+A Galaxies},} \apj, 793, 38, \dodoi{10.1088/0004-637X/793/1/38}

\bibitem[{ {Astropy Collaboration} {et~al.}(2013){Astropy Collaboration},
  {Robitaille}, {Tollerud}, {Greenfield}, {Droettboom}, {Bray}, {Aldcroft},
  {Davis}, {Ginsburg}, {Price-Whelan}, {Kerzendorf}, {Conley}, {Crighton},
  {Barbary}, {Muna}, {Ferguson}, {Grollier}, {Parikh}, {Nair}, {Unther},
  {Deil}, {Woillez}, {Conseil}, {Kramer}, {Turner}, {Singer}, {Fox}, {Weaver},
  {Zabalza}, {Edwards}, {Azalee Bostroem}, {Burke}, {Casey}, {Crawford},
  {Dencheva}, {Ely}, {Jenness}, {Labrie}, {Lim}, {Pierfederici}, {Pontzen},
  {Ptak}, {Refsdal}, {Servillat}, \& {Streicher}}]{astropy}
{Astropy Collaboration}, {Robitaille}, T.~P., {Tollerud}, E.~J., {et~al.} 2013,
  \bibinfo{title}{{Astropy: A community Python package for astronomy},} \aap,
  558, A33, \dodoi{10.1051/0004-6361/201322068}

\bibitem[{N. {Bade} {et~al.}(1996){Bade}, {Komossa}, \& {Dahlem}}]{Bade1996}
{Bade}, N., {Komossa}, S., \& {Dahlem}, M. 1996, \bibinfo{title}{{Detection of
  an extremely soft X-ray outburst in the HII-like nucleus of NGC 5905.},}
  \aap, 309, L35

\bibitem[{E.~C. {Bellm} {et~al.}(2019){Bellm}, {Kulkarni}, {Barlow}, {Feindt},
  {Graham}, {Goobar}, {Kupfer}, {Ngeow}, {Nugent}, {Ofek}, {Prince}, {Riddle},
  {Walters}, \& {Ye}}]{Bellm2019}
{Bellm}, E.~C., {Kulkarni}, S.~R., {Barlow}, T., {et~al.} 2019,
  \bibinfo{title}{{The Zwicky Transient Facility: Surveys and Scheduler},}
  \pasp, 131, 068003, \dodoi{10.1088/1538-3873/ab0c2a}

\bibitem[{N. {Blagorodnova} {et~al.}(2017){Blagorodnova}, {Gezari}, {Hung},
  {Kulkarni}, {Cenko}, {Pasham}, {Yan}, {Arcavi}, {Ben-Ami}, {Bue}, {Cantwell},
  {Cao}, {Castro-Tirado}, {Fender}, {Fremling}, {Gal-Yam}, {Ho}, {Horesh},
  {Hosseinzadeh}, {Kasliwal}, {Kong}, {Laher}, {Leloudas}, {Lunnan}, {Masci},
  {Mooley}, {Neill}, {Nugent}, {Powell}, {Valeev}, {Vreeswijk}, {Walters}, \&
  {Wozniak}}]{Blagorodnova2017}
{Blagorodnova}, N., {Gezari}, S., {Hung}, T., {et~al.} 2017,
  \bibinfo{title}{{iPTF16fnl: A Faint and Fast Tidal Disruption Event in an E+A
  Galaxy},} \apj, 844, 46, \dodoi{10.3847/1538-4357/aa7579}

\bibitem[{P.~K. {Blanchard} {et~al.}(2017){Blanchard}, {Nicholl}, {Berger},
  {Guillochon}, {Margutti}, {Chornock}, {Alexander}, {Leja}, \&
  {Drout}}]{PS16dtm}
{Blanchard}, P.~K., {Nicholl}, M., {Berger}, E., {et~al.} 2017,
  \bibinfo{title}{{PS16dtm: A Tidal Disruption Event in a Narrow-line Seyfert 1
  Galaxy},} \apj, 843, 106, \dodoi{10.3847/1538-4357/aa77f7}

\bibitem[{C.-H. {Chan} {et~al.}(2020){Chan}, {Piran}, \& {Krolik}}]{chan2020}
{Chan}, C.-H., {Piran}, T., \& {Krolik}, J.~H. 2020, \bibinfo{title}{{Light
  Curves of Tidal Disruption Events in Active Galactic Nuclei},} \apj, 903, 17,
  \dodoi{10.3847/1538-4357/abb776}

\bibitem[{C.-H. {Chan} {et~al.}(2019){Chan}, {Piran}, {Krolik}, \&
  {Saban}}]{chan2019}
{Chan}, C.-H., {Piran}, T., {Krolik}, J.~H., \& {Saban}, D. 2019,
  \bibinfo{title}{{Tidal Disruption Events in Active Galactic Nuclei},} \apj,
  881, 113, \dodoi{10.3847/1538-4357/ab2b40}

\bibitem[{T.-H. Chan {et~al.}(2015)Chan, Jia, Gao, Lu, Zeng, \& Ma}]{PCA-net1}
Chan, T.-H., Jia, K., Gao, S., {et~al.} 2015, \bibinfo{title}{PCANet: A Simple
  Deep Learning Baseline for Image Classification?} IEEE Transactions on Image
  Processing, 24, 5017, \dodoi{10.1109/TIP.2015.2475625}

\bibitem[{ {DESI Collaboration} {et~al.}(2025){DESI Collaboration},
  Abdul-Karim, Adame, Aguado, Aguilar, Ahlen, Alam, Aldering, Alexander,
  Alfarsy, Allen, Prieto, Alves, Anand, Andrade, Armengaud, Avila, Aviles,
  Awan, Bailey, Lizancos, Ballester, Bault, Bautista, BenZvi, e~Silva,
  Bermejo-Climent, Beutler, Bianchi, Blake, Blum, Bolton, Bonici, Brieden,
  Brodzeller, Brooks, Buckley-Geer, Burtin, Canning, Rosell, Carr, Carrilho,
  Casas, Castander, Cereskaite, Cervantes-Cota, Chaussidon, Chaves-Montero,
  Chen, Chen, Claybaugh, Cole, Cooper, Cousinou, Cuceu, Davis, Dawson,
  de~Belsunce, de~la Cruz, de~la Macorra, de~Mattia, Deiosso, Costa, Demina,
  Demirbozan, DeRose, Dey, Dey, Ding, Ding, Doel, Douglass, Dowicz, Ebina,
  Edelstein, Eisenstein, Elbers, Emas, Escoffier, Fagrelius, Fan, Fanning,
  Fawcett, Fernández-García, Ferraro, Findlay, Font-Ribera, Forero-Romero,
  Forero-Sánchez, Frenk, Gänsicke, Galbany, García-Bellido, Garcia-Quintero,
  Garrison, Gaztañaga, Gil-Marín, Gnedin, Gontcho, Gonzalez-Morales,
  Gonzalez-Perez, Gordon, Graur, Green, Gruen, Gsponer, Guandalin, Gutierrez,
  Guy, Hahn, Han, Han, He, Herrera-Alcantar, Honscheid, Hou, Howlett, Huterer,
  Iršič, Ishak, Jacques, Jimenez, Jing, Joachimi, Joudaki, Joyce, Jullo,
  Juneau, Karaçaylı, Karim, Kehoe, Kent, Khederlarian, Kirkby, Kisner,
  Kitaura, Kizhuprakkat, Kong, Koposov, Kremin, Krolewski, Lahav, Lai, Lamman,
  Lan, Landriau, Lang, Lange, Lasker, Goff, Guillou, Leauthaud, Levi, Li, Li,
  Lodha, Lokken, Luo, Magneville, Manera, Manser, Margala, Martini, Maus,
  McCullough, McDonald, Medina, Medina-Varela, Meisner, Mena-Fernández,
  Menegas, Mezcua, Miquel, Montero-Camacho, Moon, Moustakas, Muñoz-Gutiérrez,
  Muñoz-Santos, Myers, Myles, Nadathur, Najita, Napolitano, Newman, Nikakhtar,
  Nikutta, Niz, Noriega, Padmanabhan, Paillas, Palanque-Delabrouille, Palmese,
  Pan, Pan, Parkinson, Peacock, Percival, Pérez-Fernández, Pérez-Ràfols,
  Peterson, Piat, Pieri, Pinon, Poppett, Porredon, Prada, Pucha, Qin,
  Rabinowitz, Raichoor, Ramírez-Pérez, Ramirez-Solano, Rashkovetskyi, Ravoux,
  Riley, Rocher, Rockosi, Rohlf, Ross, Rossi, Ruggeri, Ruhlmann-Kleider, Sabiu,
  Said, Saintonge, Samushia, Sanchez, Sanders, Saulder, Schlafly, Schlegel,
  Scholte, Schubnell, Seo, Shafieloo, Sharples, Silber, Siudek, Smith,
  Sprayberry, Suárez-Pérez, Swanson, Tan, Tarlé, Taylor, Thomas, Tojeiro,
  Turner, Turner, Ureña-López, Vaisakh, Valluri, Vargas-Magaña, Verde,
  Walther, Wang, Wang, Wang, Weaver, Weaverdyck, Wechsler, White, Wolfson,
  Yang, Yèche, Youles, Yu, Yuan, Zaborowski, Zarrouk, Zhang, Zhao, Zhao,
  Zheng, Zhou, Zou, Zou, \& Zu}]{DESIDR1}
{DESI Collaboration}, Abdul-Karim, M., Adame, A.~G., {et~al.} 2025, Data
  Release 1 of the Dark Energy Spectroscopic Instrument, \doarXiv{2503.14745}

\bibitem[{J.~L. {Donley} {et~al.}(2002){Donley}, {Brandt}, {Eracleous}, \&
  {Boller}}]{Donley2002}
{Donley}, J.~L., {Brandt}, W.~N., {Eracleous}, M., \& {Boller}, T. 2002,
  \bibinfo{title}{{Large-Amplitude X-Ray Outbursts from Galactic Nuclei: A
  Systematic Survey using ROSAT Archival Data},} \aj, 124, 1308,
  \dodoi{10.1086/342280}

\bibitem[{L. {Dou} {et~al.}(2016){Dou}, {Wang}, {Jiang}, {Yang}, {Lyu}, \&
  {Zhou}}]{Dou2016}
{Dou}, L., {Wang}, T.-g., {Jiang}, N., {et~al.} 2016, \bibinfo{title}{{Long
  Fading Mid-infrared Emission in Transient Coronal Line Emitters: Dust Echo of
  a Tidal Disruption Flare},} \apj, 832, 188,
  \dodoi{10.3847/0004-637X/832/2/188}

\bibitem[{A.~J. {Drake} {et~al.}(2009){Drake}, {Djorgovski}, {Mahabal},
  {Beshore}, {Larson}, {Graham}, {Williams}, {Christensen}, {Catelan},
  {Boattini}, {Gibbs}, {Hill}, \& {Kowalski}}]{Drake2009}
{Drake}, A.~J., {Djorgovski}, S.~G., {Mahabal}, A., {et~al.} 2009,
  \bibinfo{title}{{First Results from the Catalina Real-Time Transient
  Survey},} \apj, 696, 870, \dodoi{10.1088/0004-637X/696/1/870}

\bibitem[{P. {Esquej} {et~al.}(2008){Esquej}, {Saxton}, {Komossa}, {Read},
  {Freyberg}, {Hasinger}, {Garc{\'\i}a-Hern{\'a}ndez}, {Lu}, {Rodriguez
  Zaur{\'\i}n}, {S{\'a}nchez-Portal}, \& {Zhou}}]{Esquej2008}
{Esquej}, P., {Saxton}, R.~D., {Komossa}, S., {et~al.} 2008,
  \bibinfo{title}{{Evolution of tidal disruption candidates discovered by
  XMM-Newton},} \aap, 489, 543, \dodoi{10.1051/0004-6361:200810110}

\bibitem[{E.~L. {Fitzpatrick}(1999){Fitzpatrick}}]{Fitzpatrick1999}
{Fitzpatrick}, E.~L. 1999, \bibinfo{title}{{Correcting for the Effects of
  Interstellar Extinction},} \pasp, 111, 63, \dodoi{10.1086/316293}

\bibitem[{D. {Fraix-Burnet} {et~al.}(2021){Fraix-Burnet}, {Bouveyron}, \&
  {Moultaka}}]{meansk86}
{Fraix-Burnet}, D., {Bouveyron}, C., \& {Moultaka}, J. 2021,
  \bibinfo{title}{{Unsupervised classification of SDSS galaxy spectra},} \aap,
  649, A53, \dodoi{10.1051/0004-6361/202040046}

\bibitem[{C. {Fremling} {et~al.}(2021){Fremling}, {Hall}, {Coughlin},
  {Dahiwale}, {Duev}, {Graham}, {Kasliwal}, {Kool}, {Mahabal}, {Miller},
  {Neill}, {Perley}, {Rigault}, {Rosnet}, {Rusholme}, {Sharma}, {Shin},
  {Shupe}, {Sollerman}, {Walters}, \& {Kulkarni}}]{Iascore}
{Fremling}, C., {Hall}, X.~J., {Coughlin}, M.~W., {et~al.} 2021,
  \bibinfo{title}{{SNIascore: Deep-learning Classification of Low-resolution
  Supernova Spectra},} \apjl, 917, L2, \dodoi{10.3847/2041-8213/ac116f}

\bibitem[{S. {Gezari} {et~al.}(2006){Gezari}, {Martin}, {Milliard}, {Basa},
  {Halpern}, {Forster}, {Friedman}, {Morrissey}, {Neff}, {Schiminovich},
  {Seibert}, {Small}, \& {Wyder}}]{Gezari2006}
{Gezari}, S., {Martin}, D.~C., {Milliard}, B., {et~al.} 2006,
  \bibinfo{title}{{Ultraviolet Detection of the Tidal Disruption of a Star by a
  Supermassive Black Hole},} \apjl, 653, L25, \dodoi{10.1086/509918}

\bibitem[{S. {Gezari} {et~al.}(2008){Gezari}, {Basa}, {Martin}, {Bazin},
  {Forster}, {Milliard}, {Halpern}, {Friedman}, {Morrissey}, {Neff},
  {Schiminovich}, {Seibert}, {Small}, \& {Wyder}}]{Gezari2008}
{Gezari}, S., {Basa}, S., {Martin}, D.~C., {et~al.} 2008,
  \bibinfo{title}{{UV/Optical Detections of Candidate Tidal Disruption Events
  by GALEX and CFHTLS},} \apj, 676, 944, \dodoi{10.1086/529008}

\bibitem[{S. {Gezari} {et~al.}(2009){Gezari}, {Heckman}, {Cenko}, {Eracleous},
  {Forster}, {Gon{\c{c}}alves}, {Martin}, {Morrissey}, {Neff}, {Seibert},
  {Schiminovich}, \& {Wyder}}]{Gezari2009}
{Gezari}, S., {Heckman}, T., {Cenko}, S.~B., {et~al.} 2009,
  \bibinfo{title}{{Luminous Thermal Flares from Quiescent Supermassive Black
  Holes},} \apj, 698, 1367, \dodoi{10.1088/0004-637X/698/2/1367}

\bibitem[{S. {Gezari} {et~al.}(2012){Gezari}, {Chornock}, {Rest}, {Huber},
  {Forster}, {Berger}, {Challis}, {Neill}, {Martin}, {Heckman}, {Lawrence},
  {Norman}, {Narayan}, {Foley}, {Marion}, {Scolnic}, {Chomiuk}, {Soderberg},
  {Smith}, {Kirshner}, {Riess}, {Smartt}, {Stubbs}, {Tonry}, {Wood-Vasey},
  {Burgett}, {Chambers}, {Grav}, {Heasley}, {Kaiser}, {Kudritzki}, {Magnier},
  {Morgan}, \& {Price}}]{Gezari2012}
{Gezari}, S., {Chornock}, R., {Rest}, A., {et~al.} 2012, \bibinfo{title}{{An
  ultraviolet-optical flare from the tidal disruption of a helium-rich stellar
  core},} \nat, 485, 217, \dodoi{10.1038/nature10990}

\bibitem[{S. {Goldwasser} {et~al.}(2022){Goldwasser}, {Yaron}, {Sass}, {Irani},
  {Gal-Yam}, \& {Howell}}]{NGSF}
{Goldwasser}, S., {Yaron}, O., {Sass}, A., {et~al.} 2022, \bibinfo{title}{{The
  Next Generation SuperFit (NGSF) tool is now available for online execution on
  WISeREP},} Transient Name Server AstroNote, 191, 1

\bibitem[{E. {Hammerstein} {et~al.}(2023){Hammerstein}, {van Velzen}, {Gezari},
  {Cenko}, {Yao}, {Ward}, {Frederick}, {Villanueva}, {Somalwar}, {Graham},
  {Kulkarni}, {Stern}, {Andreoni}, {Bellm}, {Dekany}, {Dhawan}, {Drake},
  {Fremling}, {Gatkine}, {Groom}, {Ho}, {Kasliwal}, {Karambelkar}, {Kool},
  {Masci}, {Medford}, {Perley}, {Purdum}, {van Roestel}, {Sharma}, {Sollerman},
  {Taggart}, \& {Yan}}]{Hammerstein2023}
{Hammerstein}, E., {van Velzen}, S., {Gezari}, S., {et~al.} 2023,
  \bibinfo{title}{{The Final Season Reimagined: 30 Tidal Disruption Events from
  the ZTF-I Survey},} \apj, 942, 9, \dodoi{10.3847/1538-4357/aca283}

\bibitem[{J.~G. {Hills}(1975){Hills}}]{Hills1975}
{Hills}, J.~G. 1975, \bibinfo{title}{{Possible power source of Seyfert galaxies
  and QSOs},} \nat, 254, 295, \dodoi{10.1038/254295a0}

\bibitem[{T.~W.-S. {Holoien} {et~al.}(2014){Holoien}, {Prieto}, {Bersier},
  {Kochanek}, {Stanek}, {Shappee}, {Grupe}, {Basu}, {Beacom}, {Brimacombe},
  {Brown}, {Davis}, {Jencson}, {Pojmanski}, \& {Szczygie{\l}}}]{Holoien2014}
{Holoien}, T.~W.-S., {Prieto}, J.~L., {Bersier}, D., {et~al.} 2014,
  \bibinfo{title}{{ASASSN-14ae: a tidal disruption event at 200 Mpc},} \mnras,
  445, 3263, \dodoi{10.1093/mnras/stu1922}

\bibitem[{T.~W.-S. {Holoien} {et~al.}(2016){Holoien}, {Kochanek}, {Prieto},
  {Stanek}, {Dong}, {Shappee}, {Grupe}, {Brown}, {Basu}, {Beacom}, {Bersier},
  {Brimacombe}, {Danilet}, {Falco}, {Guo}, {Jose}, {Herczeg}, {Long},
  {Pojmanski}, {Simonian}, {Szczygie{\l}}, {Thompson}, {Thorstensen}, {Wagner},
  \& {Wo{\'z}niak}}]{Holoien2016}
{Holoien}, T.~W.-S., {Kochanek}, C.~S., {Prieto}, J.~L., {et~al.} 2016,
  \bibinfo{title}{{Six months of multiwavelength follow-up of the tidal
  disruption candidate ASASSN-14li and implied TDE rates from ASAS-SN},}
  \mnras, 455, 2918, \dodoi{10.1093/mnras/stv2486}

\bibitem[{J.~D. Hunter(2007)Hunter}]{matplotlib}
Hunter, J.~D. 2007, \bibinfo{title}{Matplotlib: A 2D Graphics Environment,}
  Computing in Science \& Engineering, 9, 90, \dodoi{10.1109/MCSE.2007.55}

\bibitem[{N. {Jiang} {et~al.}(2016){Jiang}, {Dou}, {Wang}, {Yang}, {Lyu}, \&
  {Zhou}}]{Jiang2016}
{Jiang}, N., {Dou}, L., {Wang}, T., {et~al.} 2016, \bibinfo{title}{{The WISE
  Detection of an Infrared Echo in Tidal Disruption Event ASASSN-14li},} \apjl,
  828, L14, \dodoi{10.3847/2041-8205/828/1/L14}

\bibitem[{N. {Jiang} {et~al.}(2019){Jiang}, {Wang}, {Mou}, {Liu}, {Dou},
  {Sheng}, \& {Wang}}]{PS1-10adi}
{Jiang}, N., {Wang}, T., {Mou}, G., {et~al.} 2019, \bibinfo{title}{{Infrared
  Echo and Late-stage Rebrightening of Nuclear Transient Ps1-10adi: Exploring
  the Torus with Tidal Disruption Events in Active Galactic Nuclei},} \apj,
  871, 15, \dodoi{10.3847/1538-4357/aaf6b2}

\bibitem[{B.~C. {Kelly} {et~al.}(2009){Kelly}, {Bechtold}, \&
  {Siemiginowska}}]{Kelly2009}
{Kelly}, B.~C., {Bechtold}, J., \& {Siemiginowska}, A. 2009,
  \bibinfo{title}{{Are the Variations in Quasar Optical Flux Driven by Thermal
  Fluctuations?},} \apj, 698, 895, \dodoi{10.1088/0004-637X/698/1/895}

\bibitem[{D. Kingma \& J. Ba(2014)Kingma \& Ba}]{Adam}
Kingma, D., \& Ba, J. 2014, \bibinfo{title}{Adam: A Method for Stochastic
  Optimization,} International Conference on Learning Representations

\bibitem[{S. {Komossa} \& N. {Bade}(1999){Komossa} \& {Bade}}]{Komossa1999}
{Komossa}, S., \& {Bade}, N. 1999, \bibinfo{title}{{The giant X-ray outbursts
  in NGC 5905 and IC 3599:() hfill Follow-up observations and outburst
  scenarios},} \aap, 343, 775, \dodoi{10.48550/arXiv.astro-ph/9901141}

\bibitem[{S. Lanthaler(2023)Lanthaler}]{PCA-Net2}
Lanthaler, S. 2023, \bibinfo{title}{Operator learning with PCA-Net: upper and
  lower complexity bounds,} Journal of Machine Learning Research, 24, 1.
\newblock \url{http://jmlr.org/papers/v24/23-0478.html}

\bibitem[{G.~R. Lee {et~al.}(2019)Lee, Gommers, Waselewski, Wohlfahrt, \&
  {O'Leary}}]{Lee2019}
Lee, G.~R., Gommers, R., Waselewski, F., Wohlfahrt, K., \& {O'Leary}, A. 2019,
  \bibinfo{title}{PyWavelets: A Python package for wavelet analysis,} Journal
  of Open Source Software, 4, 1237, \dodoi{10.21105/joss.01237}

\bibitem[{A.~P. {Lightman} \& S.~L. {Shapiro}(1977){Lightman} \&
  {Shapiro}}]{Lightman1977}
{Lightman}, A.~P., \& {Shapiro}, S.~L. 1977, \bibinfo{title}{{The distribution
  and consumption rate of stars around a massive, collapsed object.},} \apj,
  211, 244, \dodoi{10.1086/154925}

\bibitem[{Z. {Lin} {et~al.}(2025){Lin}, {Jiang}, {Wang}, {Kong}, {Huang},
  {Lin}, {Qin}, \& {Xia}}]{2022fpx}
{Lin}, Z., {Jiang}, N., {Wang}, Y., {et~al.} 2025, \bibinfo{title}{{Insights
  from the ``Red Devil'' AT 2022fpx: A Dust-reddened Family of Tidal Disruption
  Events Excluded by Their Apparent Red Color?},} \apj, 990, 22,
  \dodoi{10.3847/1538-4357/adef10}

\bibitem[{Z. {Liu} {et~al.}(2020){Liu}, {Li}, {Liu}, {Lu}, {Yuan}, {Dou}, \&
  {Shen}}]{Liu2020}
{Liu}, Z., {Li}, D., {Liu}, H.-Y., {et~al.} 2020, \bibinfo{title}{{A Tidal
  Disruption Event Candidate Discovered in the Active Galactic Nucleus SDSS
  J022700.77-042020.6},} \apj, 894, 93, \dodoi{10.3847/1538-4357/ab880f}

\bibitem[{C.~L. {MacLeod} {et~al.}(2010){MacLeod}, {Ivezi{\'c}}, {Kochanek},
  {Koz{\l}owski}, {Kelly}, {Bullock}, {Kimball}, {Sesar}, {Westman}, {Brooks},
  {Gibson}, {Becker}, \& {de Vries}}]{MacLeod2010}
{MacLeod}, C.~L., {Ivezi{\'c}}, {\v{Z}}., {Kochanek}, C.~S., {et~al.} 2010,
  \bibinfo{title}{{Modeling the Time Variability of SDSS Stripe 82 Quasars as a
  Damped Random Walk},} \apj, 721, 1014, \dodoi{10.1088/0004-637X/721/2/1014}

\bibitem[{J. {Magorrian} \& S. {Tremaine}(1999){Magorrian} \&
  {Tremaine}}]{Magorrian1999}
{Magorrian}, J., \& {Tremaine}, S. 1999, \bibinfo{title}{{Rates of tidal
  disruption of stars by massive central black holes},} \mnras, 309, 447,
  \dodoi{10.1046/j.1365-8711.1999.02853.x}

\bibitem[{D.~C. {Martin} {et~al.}(2005){Martin}, {Fanson}, {Schiminovich},
  {Morrissey}, {Friedman}, {Barlow}, {Conrow}, {Grange}, {Jelinsky},
  {Milliard}, {Siegmund}, {Bianchi}, {Byun}, {Donas}, {Forster}, {Heckman},
  {Lee}, {Madore}, {Malina}, {Neff}, {Rich}, {Small}, {Surber}, {Szalay},
  {Welsh}, \& {Wyder}}]{Martin2005}
{Martin}, D.~C., {Fanson}, J., {Schiminovich}, D., {et~al.} 2005,
  \bibinfo{title}{{The Galaxy Evolution Explorer: A Space Ultraviolet Survey
  Mission},} \apjl, 619, L1, \dodoi{10.1086/426387}

\bibitem[{P. {Mil{\'a}n Veres} {et~al.}(2024){Mil{\'a}n Veres}, {Franckowiak},
  {van Velzen}, {Adebahr}, {Taziaux}, {Necker}, {Stein}, {Kier}, {Mueller},
  {Bomans}, {Jordana-Mitjans}, {Kowalski}, {Hammerstein}, {Marci-Boehncke},
  {Reusch}, {Garrappa}, {Rose}, \& {Kashyap Das}}]{2019aalc}
{Mil{\'a}n Veres}, P., {Franckowiak}, A., {van Velzen}, S., {et~al.} 2024,
  \bibinfo{title}{{Back from the dead: AT2019aalc as a candidate repeating TDE
  in an AGN},} arXiv e-prints, arXiv:2408.17419,
  \dodoi{10.48550/arXiv.2408.17419}

\bibitem[{J.~B. {Oke}(1974){Oke}}]{Oke1974}
{Oke}, J.~B. 1974, \bibinfo{title}{{Absolute Spectral Energy Distributions for
  White Dwarfs},} \apjs, 27, 21, \dodoi{10.1086/190287}

\bibitem[{D.~E. {Osterbrock} \& G.~J. {Ferland}(2006){Osterbrock} \&
  {Ferland}}]{Osterbrock2006}
{Osterbrock}, D.~E., \& {Ferland}, G.~J. 2006, {Astrophysics of gaseous nebulae
  and active galactic nuclei} (University Science Books)

\bibitem[{A. Paszke {et~al.}(2019)Paszke, Gross, Massa, Lerer, Bradbury,
  Chanan, Killeen, Lin, Gimelshein, Antiga, Desmaison, Kopf, Yang, DeVito,
  Raison, Tejani, Chilamkurthy, Steiner, Fang, Bai, \& Chintala}]{pytorch2}
Paszke, A., Gross, S., Massa, F., {et~al.} 2019, \bibinfo{title}{PyTorch: An
  Imperative Style, High-Performance Deep Learning Library,} in Advances in
  Neural Information Processing Systems 32, ed. H.~Wallach, H.~Larochelle,
  A.~Beygelzimer, F.~d\textquotesingle Alch\'{e}-Buc, E.~Fox, \& R.~Garnett
  (Curran Associates, Inc.), 8024--8035.
\newblock \url{http://papers.neurips.cc/paper/9015-pytorch-an \
  -imperative-style-high-performance-deep-learning-library.pdf}

\bibitem[{ {Planck Collaboration} {et~al.}(2016){Planck Collaboration},
  {Aghanim}, {Ashdown}, {Aumont}, {Baccigalupi}, {Ballardini}, {Banday},
  {Barreiro}, {Bartolo}, {Basak}, {Benabed}, {Bernard}, {Bersanelli},
  {Bielewicz}, {Bonavera}, {Bond}, {Borrill}, {Bouchet}, {Boulanger},
  {Burigana}, {Calabrese}, {Cardoso}, {Carron}, {Chiang}, {Colombo}, {Comis},
  {Couchot}, {Coulais}, {Crill}, {Curto}, {Cuttaia}, {de Bernardis}, {de
  Zotti}, {Delabrouille}, {Di Valentino}, {Dickinson}, {Diego}, {Dor{\'e}},
  {Douspis}, {Ducout}, {Dupac}, {Dusini}, {Elsner}, {En{\ss}lin}, {Eriksen},
  {Falgarone}, {Fantaye}, {Finelli}, {Forastieri}, {Frailis}, {Fraisse},
  {Franceschi}, {Frolov}, {Galeotta}, {Galli}, {Ganga}, {G{\'e}nova-Santos},
  {Gerbino}, {Ghosh}, {Giraud-H{\'e}raud}, {Gonz{\'a}lez-Nuevo}, {G{\'o}rski},
  {Gruppuso}, {Gudmundsson}, {Hansen}, {Helou}, {Henrot-Versill{\'e}},
  {Herranz}, {Hivon}, {Huang}, {Jaffe}, {Jones}, {Keih{\"a}nen}, {Keskitalo},
  {Kiiveri}, {Kisner}, {Krachmalnicoff}, {Kunz}, {Kurki-Suonio}, {Lamarre},
  {Langer}, {Lasenby}, {Lattanzi}, {Lawrence}, {Le Jeune}, {Levrier}, {Lilje},
  {Lilley}, {Lindholm}, {L{\'o}pez-Caniego}, {Ma}, {Mac{\'\i}as-P{\'e}rez},
  {Maggio}, {Maino}, {Mandolesi}, {Mangilli}, {Maris}, {Martin},
  {Mart{\'\i}nez-Gonz{\'a}lez}, {Matarrese}, {Mauri}, {McEwen}, {Melchiorri},
  {Mennella}, {Migliaccio}, {Miville-Desch{\^e}nes}, {Molinari}, {Moneti},
  {Montier}, {Morgante}, {Moss}, {Natoli}, {Oxborrow}, {Pagano}, {Paoletti},
  {Patanchon}, {Perdereau}, {Perotto}, {Pettorino}, {Piacentini},
  {Plaszczynski}, {Polastri}, {Polenta}, {Puget}, {Rachen}, {Racine},
  {Reinecke}, {Remazeilles}, {Renzi}, {Rocha}, {Rosset}, {Rossetti}, {Roudier},
  {Rubi{\~n}o-Mart{\'\i}n}, {Ruiz-Granados}, {Salvati}, {Sandri}, {Savelainen},
  {Scott}, {Sirignano}, {Sirri}, {Soler}, {Spencer}, {Suur-Uski}, {Tauber},
  {Tavagnacco}, {Tenti}, {Toffolatti}, {Tomasi}, {Tristram}, {Trombetti},
  {Valiviita}, {Van Tent}, {Vielva}, {Villa}, {Vittorio}, {Wandelt}, {Wehus},
  {Zacchei}, \& {Zonca}}]{Planck2016}
{Planck Collaboration}, {Aghanim}, N., {Ashdown}, M., {et~al.} 2016,
  \bibinfo{title}{{Planck intermediate results. XLVIII. Disentangling Galactic
  dust emission and cosmic infrared background anisotropies},} \aap, 596, A109,
  \dodoi{10.1051/0004-6361/201629022}

\bibitem[{M.~J. {Rees}(1988){Rees}}]{Rees1988}
{Rees}, M.~J. 1988, \bibinfo{title}{{Tidal disruption of stars by black holes
  of {}10$^{6}$-{}10$^{8}$ solar masses in nearby galaxies},} \nat, 333, 523,
  \dodoi{10.1038/333523a0}

\bibitem[{M. {St. Clair} \& C. {Million}(2022){St. Clair} \&
  {Million}}]{Stclair2022}
{St. Clair}, M., \& {Million}, C. 2022, {MillionConcepts/gPhoton2: v3.0.0a0},
  v3.0.0a0 Zenodo, \dodoi{10.5281/zenodo.7472061}

\bibitem[{D. {Stern} {et~al.}(2012){Stern}, {Assef}, {Benford}, {Blain},
  {Cutri}, {Dey}, {Eisenhardt}, {Griffith}, {Jarrett}, {Lake}, {Masci},
  {Petty}, {Stanford}, {Tsai}, {Wright}, {Yan}, {Harrison}, \&
  {Madsen}}]{Stern2012}
{Stern}, D., {Assef}, R.~J., {Benford}, D.~J., {et~al.} 2012,
  \bibinfo{title}{{Mid-infrared Selection of Active Galactic Nuclei with the
  Wide-Field Infrared Survey Explorer. I. Characterizing WISE-selected Active
  Galactic Nuclei in COSMOS},} \apj, 753, 30,
  \dodoi{10.1088/0004-637X/753/1/30}

\bibitem[{N.~C. {Stone} \& B.~D. {Metzger}(2016){Stone} \&
  {Metzger}}]{Stone2016}
{Stone}, N.~C., \& {Metzger}, B.~D. 2016, \bibinfo{title}{{Rates of stellar
  tidal disruption as probes of the supermassive black hole mass function},}
  \mnras, 455, 859, \dodoi{10.1093/mnras/stv2281}

\bibitem[{D. {Syer} \& A. {Ulmer}(1999){Syer} \& {Ulmer}}]{Syer1999}
{Syer}, D., \& {Ulmer}, A. 1999, \bibinfo{title}{{Tidal disruption rates of
  stars in observed galaxies},} \mnras, 306, 35,
  \dodoi{10.1046/j.1365-8711.1999.02445.x}

\bibitem[{S. {van Velzen}(2018){van Velzen}}]{vanVelzen2018}
{van Velzen}, S. 2018, \bibinfo{title}{{On the Mass and Luminosity Functions of
  Tidal Disruption Flares: Rate Suppression due to Black Hole Event Horizons},}
  \apj, 852, 72, \dodoi{10.3847/1538-4357/aa998e}

\bibitem[{S. {van Velzen} \& G.~R. {Farrar}(2014){van Velzen} \&
  {Farrar}}]{vanVelzen2014}
{van Velzen}, S., \& {Farrar}, G.~R. 2014, \bibinfo{title}{{Measurement of the
  Rate of Stellar Tidal Disruption Flares},} \apj, 792, 53,
  \dodoi{10.1088/0004-637X/792/1/53}

\bibitem[{S. {van Velzen} {et~al.}(2016){van Velzen}, {Mendez}, {Krolik}, \&
  {Gorjian}}]{vanVelzen2016}
{van Velzen}, S., {Mendez}, A.~J., {Krolik}, J.~H., \& {Gorjian}, V. 2016,
  \bibinfo{title}{{Discovery of Transient Infrared Emission from Dust Heated by
  Stellar Tidal Disruption Flares},} \apj, 829, 19,
  \dodoi{10.3847/0004-637X/829/1/19}

\bibitem[{S. {van Velzen} {et~al.}(2011){van Velzen}, {Farrar}, {Gezari},
  {Morrell}, {Zaritsky}, {{\"O}stman}, {Smith}, {Gelfand}, \&
  {Drake}}]{vanVelzen2011}
{van Velzen}, S., {Farrar}, G.~R., {Gezari}, S., {et~al.} 2011,
  \bibinfo{title}{{Optical Discovery of Probable Stellar Tidal Disruption
  Flares},} \apj, 741, 73, \dodoi{10.1088/0004-637X/741/2/73}

\bibitem[{S. {van Velzen} {et~al.}(2021{\natexlab{a}}){van Velzen}, {Gezari},
  {Hammerstein}, {Roth}, {Frederick}, {Ward}, {Hung}, {Cenko}, {Stein},
  {Perley}, {Taggart}, {Foley}, {Sollerman}, {Blagorodnova}, {Andreoni},
  {Bellm}, {Brinnel}, {De}, {Dekany}, {Feeney}, {Fremling}, {Giomi}, {Golkhou},
  {Graham}, {Ho}, {Kasliwal}, {Kilpatrick}, {Kulkarni}, {Kupfer}, {Laher},
  {Mahabal}, {Masci}, {Miller}, {Nordin}, {Riddle}, {Rusholme}, {van Santen},
  {Sharma}, {Shupe}, \& {Soumagnac}}]{vanVelzen2021}
{van Velzen}, S., {Gezari}, S., {Hammerstein}, E., {et~al.} 2021{\natexlab{a}},
  \bibinfo{title}{{Seventeen Tidal Disruption Events from the First Half of ZTF
  Survey Observations: Entering a New Era of Population Studies},} \apj, 908,
  4, \dodoi{10.3847/1538-4357/abc258}

\bibitem[{S. {van Velzen} {et~al.}(2021{\natexlab{b}}){van Velzen}, {Gezari},
  {Hammerstein}, {Roth}, {Frederick}, {Ward}, {Hung}, {Cenko}, {Stein},
  {Perley}, {Taggart}, {Foley}, {Sollerman}, {Blagorodnova}, {Andreoni},
  {Bellm}, {Brinnel}, {De}, {Dekany}, {Feeney}, {Fremling}, {Giomi}, {Golkhou},
  {Graham}, {Ho}, {Kasliwal}, {Kilpatrick}, {Kulkarni}, {Kupfer}, {Laher},
  {Mahabal}, {Masci}, {Miller}, {Nordin}, {Riddle}, {Rusholme}, {van Santen},
  {Sharma}, {Shupe}, \& {Soumagnac}}]{17TDE}
{van Velzen}, S., {Gezari}, S., {Hammerstein}, E., {et~al.} 2021{\natexlab{b}},
  \bibinfo{title}{{Seventeen Tidal Disruption Events from the First Half of ZTF
  Survey Observations: Entering a New Era of Population Studies},} \apj, 908,
  4, \dodoi{10.3847/1538-4357/abc258}

\bibitem[{P. Virtanen {et~al.}(2020)Virtanen, Gommers, Oliphant, Haberland,
  Reddy, Cournapeau, Burovski, Peterson, Weckesser, Bright, van~der Walt,
  Brett, Wilson, Millman, Mayorov, Nelson, Jones, Kern, Larson, Carey, Polat,
  Feng, Moore, VanderPlas, Laxalde, Perktold, Cimrman, Henriksen, Quintero,
  Harris, Archibald, Ribeiro, Pedregosa, van Mulbregt, Vijaykumar, Bardelli,
  Rothberg, Hilboll, Kloeckner, Scopatz, Lee, Rokem, Woods, Fulton, Masson,
  Häggström, Fitzgerald, Nicholson, Hagen, Pasechnik, Olivetti, Martin,
  Wieser, Silva, Lenders, Wilhelm, Young, Price, Ingold, Allen, Lee, Audren,
  Probst, Dietrich, Silterra, Webber, Slavič, Nothman, Buchner, Kulick,
  Schönberger, de~Miranda~Cardoso, Reimer, Harrington, Rodríguez,
  Nunez-Iglesias, Kuczynski, Tritz, Thoma, Newville, Kümmerer, Bolingbroke,
  Tartre, Pak, Smith, Nowaczyk, Shebanov, Pavlyk, Brodtkorb, Lee, McGibbon,
  Feldbauer, Lewis, Tygier, Sievert, Vigna, Peterson, More, Pudlik, Oshima,
  {et~al.}}]{scipy}
Virtanen, P., Gommers, R., Oliphant, T.~E., {et~al.} 2020,
  \bibinfo{title}{SciPy 1.0: fundamental algorithms for scientific computing in
  Python,} Nature Methods, 17, 261, \dodoi{10.1038/s41592-019-0686-2}

\bibitem[{J. {Wang} \& D. {Merritt}(2004){Wang} \& {Merritt}}]{Wang2004}
{Wang}, J., \& {Merritt}, D. 2004, \bibinfo{title}{{Revised Rates of Stellar
  Disruption in Galactic Nuclei},} \apj, 600, 149, \dodoi{10.1086/379767}

\bibitem[{T.-G. {Wang} {et~al.}(2012){Wang}, {Zhou}, {Komossa}, {Wang}, {Yuan},
  \& {Yang}}]{Wang2012}
{Wang}, T.-G., {Zhou}, H.-Y., {Komossa}, S., {et~al.} 2012,
  \bibinfo{title}{{Extreme Coronal Line Emitters: Tidal Disruption of Stars by
  Massive Black Holes in Galactic Nuclei?},} \apj, 749, 115,
  \dodoi{10.1088/0004-637X/749/2/115}

\bibitem[{J. Wen {et~al.}(2024)Wen, Zhang, Lu, Hu, \& Huang}]{Wen2024}
Wen, J., Zhang, N., Lu, X., Hu, Z., \& Huang, H. 2024,
  \bibinfo{title}{Mgformer: Multi-group transformer for multivariate time
  series classification,} Engineering Applications of Artificial Intelligence,
  133, 108633, \dodoi{https://doi.org/10.1016/j.engappai.2024.108633}

\bibitem[{E.~L. {Wright} {et~al.}(2010){Wright}, {Eisenhardt}, {Mainzer},
  {Ressler}, {Cutri}, {Jarrett}, {Kirkpatrick}, {Padgett}, {McMillan},
  {Skrutskie}, {Stanford}, {Cohen}, {Walker}, {Mather}, {Leisawitz}, {Gautier},
  {McLean}, {Benford}, {Lonsdale}, {Blain}, {Mendez}, {Irace}, {Duval}, {Liu},
  {Royer}, {Heinrichsen}, {Howard}, {Shannon}, {Kendall}, {Walsh}, {Larsen},
  {Cardon}, {Schick}, {Schwalm}, {Abid}, {Fabinsky}, {Naes}, \&
  {Tsai}}]{Wright2010}
{Wright}, E.~L., {Eisenhardt}, P. R.~M., {Mainzer}, A.~K., {et~al.} 2010,
  \bibinfo{title}{{The Wide-field Infrared Survey Explorer (WISE): Mission
  Description and Initial On-orbit Performance},} \aj, 140, 1868.
\newblock \doarXiv{1008.0031}

\bibitem[{Y. {Yao} {et~al.}(2023){Yao}, {Ravi}, {Gezari}, {van Velzen}, {Lu},
  {Schulze}, {Somalwar}, {Kulkarni}, {Hammerstein}, {Nicholl}, {Graham},
  {Perley}, {Cenko}, {Stein}, {Ricarte}, {Chadayammuri}, {Quataert}, {Bellm},
  {Bloom}, {Dekany}, {Drake}, {Groom}, {Mahabal}, {Prince}, {Riddle},
  {Rusholme}, {Sharma}, {Sollerman}, \& {Yan}}]{Yao2023}
{Yao}, Y., {Ravi}, V., {Gezari}, S., {et~al.} 2023, \bibinfo{title}{{Tidal
  Disruption Event Demographics with the Zwicky Transient Facility: Volumetric
  Rates, Luminosity Function, and Implications for the Local Black Hole Mass
  Function},} \apjl, 955, L6, \dodoi{10.3847/2041-8213/acf216}

\bibitem[{O. {Yaron} \& A. {Gal-Yam}(2012){Yaron} \& {Gal-Yam}}]{Yaron2012}
{Yaron}, O., \& {Gal-Yam}, A. 2012, \bibinfo{title}{{WISeREP{\textemdash}An
  Interactive Supernova Data Repository},} \pasp, 124, 668,
  \dodoi{10.1086/666656}

\bibitem[{A. {Zabludoff} {et~al.}(2021){Zabludoff}, {Arcavi}, {LaMassa},
  {Perets}, {Trakhtenbrot}, {Zauderer}, {Auchettl}, {Dai}, {French}, {Hung},
  {Kara}, {Lodato}, {Maksym}, {Qin}, {Ramirez-Ruiz}, {Roth}, {Runnoe}, \&
  {Wevers}}]{distinguish_TDE}
{Zabludoff}, A., {Arcavi}, I., {LaMassa}, S., {et~al.} 2021,
  \bibinfo{title}{{Distinguishing Tidal Disruption Events from Impostors},}
  \ssr, 217, 54, \dodoi{10.1007/s11214-021-00829-4}

\bibitem[{R. {Zheng} {et~al.}(2026){Zheng}, {Lin}, {Kong}, {Meng}, {Xu}, {Fan},
  {Jiang}, {Jiang}, {Lin}, {Wang}, {Zhu}, {Li}, {Liang}, {Liu}, {Lou}, {Luo},
  {Tang}, {Wang}, {Wang}, {Xue}, {Yao}, {Zhang}, {Zhao}, {Zheng}, \&
  {Zuo}}]{Zheng2025}
{Zheng}, R., {Lin}, Z., {Kong}, X., {et~al.} 2026, \bibinfo{title}{{TTC:
  Transformer-based TDE Classifier for the Wide Field Survey Telescope
  (WFST)},} \apj, 999, 181, \dodoi{10.3847/1538-4357/ae3153}

\end{thebibliography}
\bibliographystyle{aasjournalv7}



\clearpage

\appendix











\section{Supplementary figures and tables for classification of model-selected TDE candidates}

\begin{figure*}[h]
\centering
\figurenum{A1}
\includegraphics[width=0.9\textwidth]{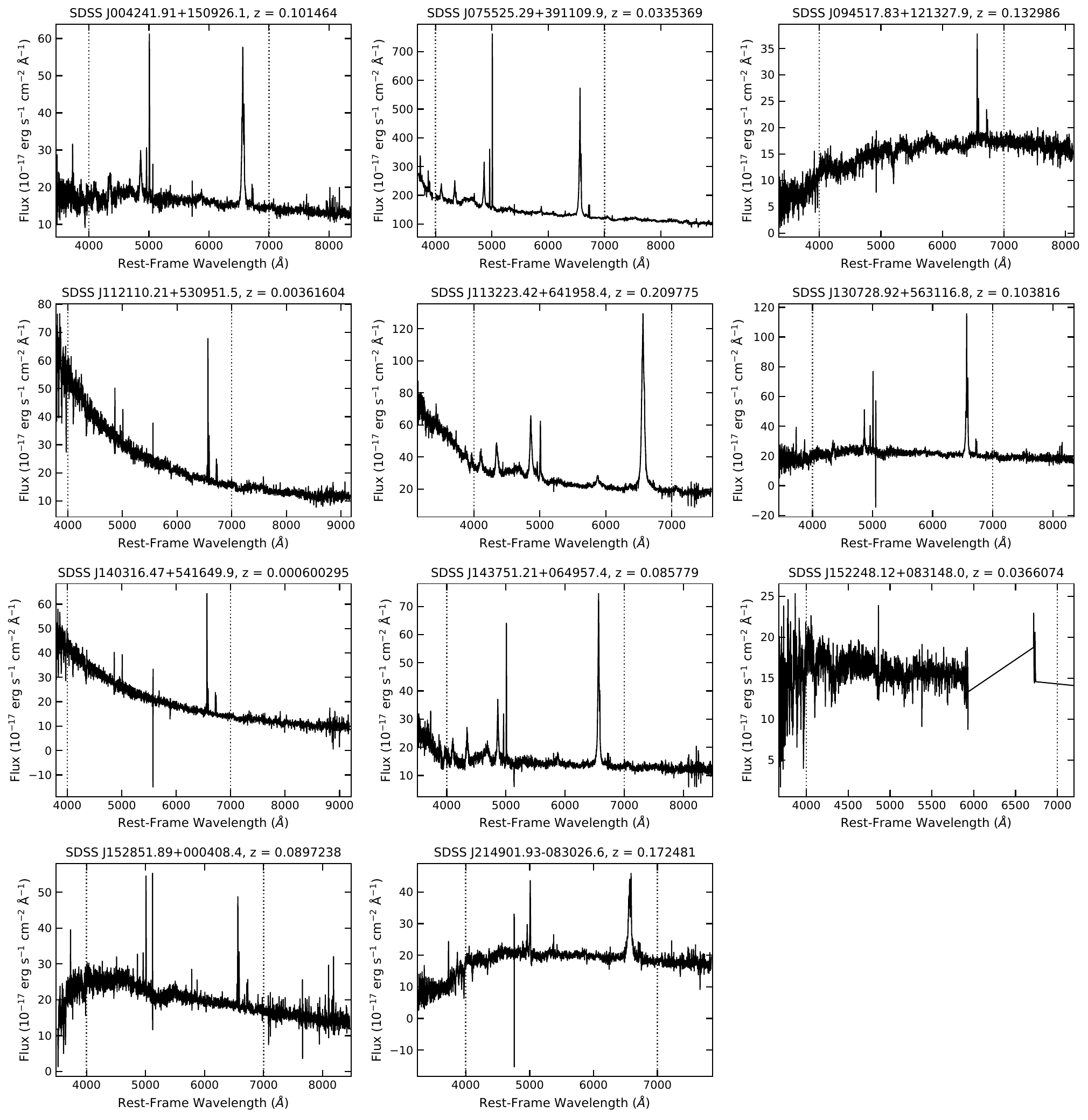}
\caption{Spectra for 11 model-selected TDE candidates that were subsequently confirmed to be not TDEs or unlikely to be TDEs. Black dotted vertical lines indicate the wavelength range from 4000 to 7000~\AA, which was the range that extracted and used for the classification in our work.
\label{fig:bogus}}
\end{figure*}

\begin{figure*}[h]
\centering
\figurenum{A2}
\includegraphics[width=0.75\textwidth]{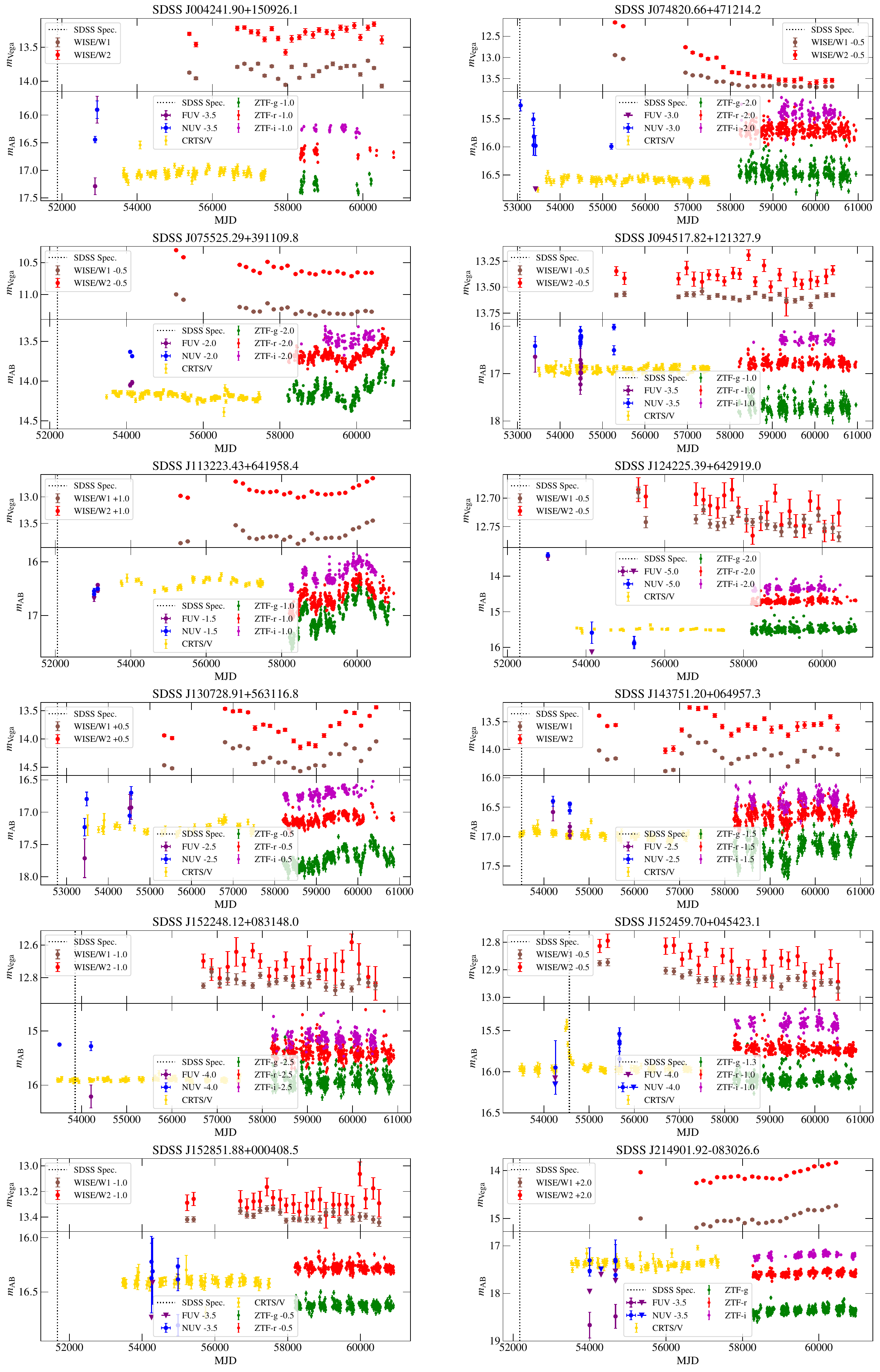}
\caption{UV (GALEX), optical (CRTS \& ZTF) and mid-infrared (WISE) light curves for 12 TDE candidates with $z>0.005$. SDSS J124225.39+642919.0 and SDSS J152459.70+045423.1 are two newly discovered TDEs, SDSS J074820.66+471214.2 is a likely TDE already reported by \citet{Wang2012}. 
\label{fig:uv_opt_ir_lc}}
\end{figure*}

This supplementary material provides additional figures and tables for the TDE candidates utilized in this study. Figure~\ref{fig:bogus} illustrates the original rest-frame spectra of sources identified as non-TDEs or deemed improbable candidates following secondary screening. Figure~\ref{fig:uv_opt_ir_lc} presents the multi-band photometric light curves for all candidates situated at redshifts $z > 0.005$.


\section{Statistical Interpretation of TDE Scores and Classification Confidence}\label{sec:score_discussion}

\begin{figure*}[h]
\centering
\figurenum{B1}
\includegraphics[width=0.9\textwidth]{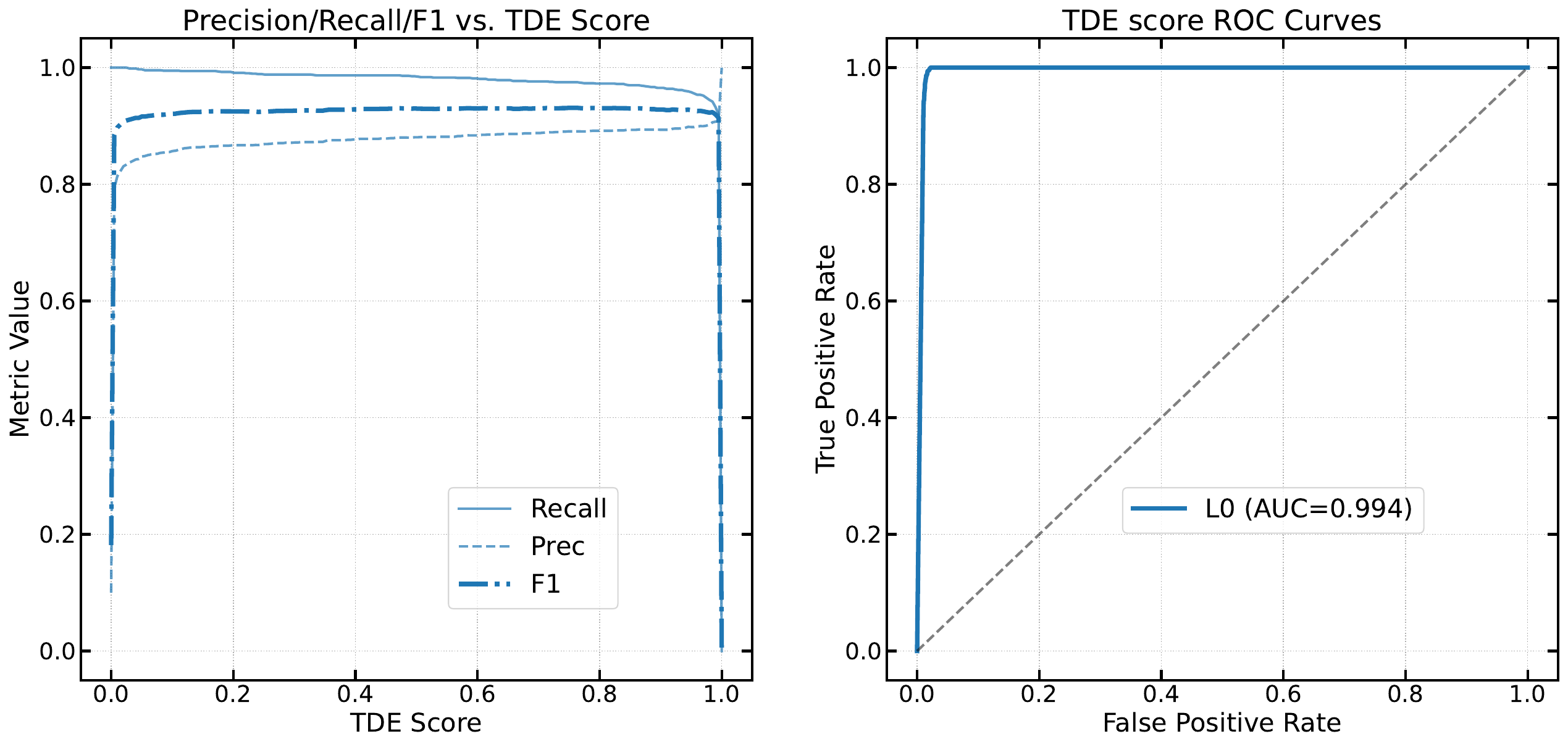}
\caption{Statistical performance evaluation of the TDE classification pipeline. Left panel: variations of Precision, Recall, and F1-score as functions of the TDE Score threshold. Right panel: ROC curve plotting TPR against FPR. An AUC value approaching unity signifies superior classification efficacy and heightened discriminatory power of the architecture.
\label{fig:me-mix}}
\end{figure*}

In Figure~\ref{fig:me-mix}, we evaluate the model’s performance on the test set by analyzing the variations of Precision, Recall, and the F1-score as functions of the TDE Score threshold, alongside the Receiver Operating Characteristic (ROC) curve.

Precision represents the "purity" of the classified sample, indicating the fraction of true TDEs among all sources identified as such by the model. Recall (or completeness) measures the fraction of actual TDEs in the dataset that the model successfully recovers. F1 score provides a harmonic mean of the two, serving as a balanced metric for overall classification efficacy.

The ROC curve illustrates the diagnostic ability of the classifier by plotting the True Positive Rate against the False Positive Rate; the Area Under the Curve (AUC) quantifies the model's ability to distinguish between TDEs and non-TDE transients. The performance of the model is positively correlated with the AUC value; an AUC approaching unity signifies superior classification power and higher diagnostic accuracy.

Ideally, a physical TDE source should yield a TDE Score approaching unity. However, these probabilistic outputs must be interpreted with caution.
The interpretive reliability of the classification scores generated by our pipeline is primarily influenced by two distinct factors.      

First, the current architecture relies predominantly on PCA-derived components for spectral representation, which inherently omits high-order spectral features contained within the PCA residuals. Since the principal eigenvectors do not capture these residuals, crucial diagnostic information, such as narrow emission lines essential for distinguishing TDEs from other transients is not fully accounted for by the network. Consequently, sources with similar PCA coefficients but distinct physical morphologies may be assigned similar scores, leading to the absence of a sharp distribution gap between high-confidence TDEs and "unlikely" candidates in real data. While incorporating these residuals is a logical progression, they are often dominated by stochastic noise, and their high dimensionality presents significant challenges for compression via standard linear methods without introducing excessive degrees of freedom.      

Second, the model performance is inherently constrained by the limited diversity and sample size of currently available TDE templates. Our training set primarily utilizes spectra from the WISeREP database; however, the absolute number of "standard" TDE spectra archived in such repositories remains remarkably scarce. This scarcity, combined with the lack of a specific sub-classification for “Featureless-TDE” events sources that exhibit a blue continuum similar to typical TDEs but lack prominent emission lines further limits the model's ability to generalize across the full TDE population. Therefore, a lower classification score does not definitively exclude a source from being a TDE; rather, it may signify that the target deviates from the standard archetypes present in the sparse training library. This uncertainty reflects a lack of representativeness in current public spectral repositories rather than a fundamental failure of the model architecture itself.

\end{document}